\documentclass[twocolumn]{openjournal}

\usepackage{siunitx}
\usepackage{amsmath}
\usepackage{longtable}
\usepackage{graphicx}
\usepackage{booktabs}
\usepackage{appendix}
\usepackage{url}
\usepackage{hyperref}

\graphicspath{{FIGURES/}}
\shortauthors{Clancy et al.}
\shorttitle{Transient Detection Pipeline for SO LAT TOD}

\begin{document}

\title{The Simons Observatory: Development of a Pipeline to Detect Rapid Transients in Time-Ordered Data}


\author{
    \href{http://orcid.org/0000-0002-9711-9969}{Justin~Clancy$^{\star,1}$},
    \href{http://orcid.org/0000-0002-8211-1630}{Carlo~Baccigalupi$^{2,3,4,5}$},
    \href{http://orcid.org/0000-0002-1327-1921}{Josh~Borrow$^{6}$},
    \href{http://orcid.org/0009-0006-7382-1434}{Nadia~Dachlythra$^{7}$},
    \href{http://orcid.org/0000-0002-7145-1824}{Allen~M.~Foster$^{8}$},
    \href{http://orcid.org/0000-0002-1697-3080}{Yilun~Guan$^{9}$},
    \href{http://orcid.org/0000-0002-4765-3426}{Carlos~Herv\'ias-Caimapo$^{10}$},
    \href{http://orcid.org/0000-0003-1690-6678}{Adam~D.~Hincks$^{9,11}$},
    \href{http://orcid.org/0000-0002-0965-7864}{Renee~Hlozek$^{12,9}$},
    \href{http://orcid.org/0009-0004-8314-2043}{Erika~Hornecker$^{9}$},
    \href{http://orcid.org/0000-0001-7109-0099}{Kevin~M.~Huffenberger$^{13}$},
    \href{http://orcid.org/0000-0001-5748-5182}{Reijo~Keskitalo$^{14,15}$},
    \href{http://orcid.org/0009-0006-0076-2613}{Simran~K.~Nerval$^{9,12}$},
    \href{http://orcid.org/0000-0003-1842-8104}{John~Orlowski-Scherer$^{6}$},
    \href{http://orcid.org/0000-0003-3412-2586}{Mike~W.~Peel$^{16}$},
    \href{http://orcid.org/0000-0003-2226-9169}{Christian~L.~Reichardt$^{1}$},
    \href{http://orcid.org/0000-0001-5327-1400}{Cristian~Vargas$^{13}$}
}
\email[$^\star$]{clancyj1@student.unimelb.edu.au}

\affiliation{
$^1$School of Physics, The University of Melbourne, Parkville VIC 3010, Australia\\
$^2$The International School for Advanced Studies (SISSA), via Bonomea 265, I-34136 Trieste, Italy\\
$^3$The National Institute for Nuclear Physics (INFN), via Valerio 2, I-34127, Trieste, Italy\\
$^4$The National Institute for Astrophysics (INAF), via Tiepolo 11, I-34143, Trieste, Italy\\
$^5$The Institute for Fundamental Physics of the Universe (IFPU), Via Beirut 2, I-34151, Trieste, Italy\\
$^6$Department of Physics and Astronomy, University of Pennsylvania, 209 South 33rd Street, Philadelphia, PA 19104, USA\\
$^7$Department of Physics, University of Milano-Bicocca, Piazza della Scienza 3, 20126 Milano (MI), Italy\\
$^8$Department of Physics, Princeton University - Jadwin Hall, Princeton University, Princeton, NJ 08544, USA\\
$^{9}$David A. Dunlap Department of Astronomy \& Astrophysics, University of Toronto, 50 St. George Street, Toronto, ON, M5S 3H4, Canada\\
$^{10}$Instituto de Astrof\'isica and Centro de Astro-Ingenier\'ia, Facultad de F\'isica, Pontificia Universidad Cat\'olica de Chile, Av. Vicu\~na Mackenna 4860, 7820436 Macul, Santiago, Chile\\
$^{11}$Specola Vaticana (Vatican Observatory), V-00120 Vatican City State\\
$^{12}$Dunlap Institute for Astronomy \& Astrophysics, University of Toronto, 50 St. George Street, Toronto, ON, M5S 3H4, Canada\\
$^{13}$Mitchell Institute for Fundamental Physics \& Astronomy, Department of Physics \& Astronomy, Texas A\&M University, College Station, Texas, 77843 USA\\
$^{14}$University of Oslo, Institute of Theoretical Astrophysics, Oslo, Norway\\
$^{15}$Lawrence Berkeley National Laboratory, Berkeley, CA, USA\\
$^{16}$Imperial College London, Blackett Lab, Prince Consort Road, London SW7 2AZ, UK\\
}

\begin{abstract}
We introduce a method for detecting astrophysical transients evolving on timescales of milliseconds to minutes using cosmic microwave background (CMB) survey telescopes.
While previous transient searches in CMB data operate in map space, our pipeline directly processes the raw time-ordered data, enabling sensitivity to fast, dynamic signals.
We integrate our detection approach into the Simons Observatory time-domain pipeline and assess the performance by injecting symmetric, stellar flare-like light curves into simulated observations.
For events flaring with a timescale of 0.5\,s, the pipeline detects $\gtrsim90$\% of events at flux densities of 800, 1150, 1650, and 4250\,mJy when measured in the 93, 145, 225, and 280\,GHz bands respectively.
At a fixed peak flux density, the pipeline more readily detects longer flares.
The limiting flux density for 90\% completeness is four times lower for a $\ge5$\,s flare than for a 0.5\,s flare, while the flux density limits for $\gtrsim50$\% detection efficiency are comparable to the rms noise of the time-ordered data.
We are able to determine the position of detected events in each observing band, with a positional uncertainty at the detection threshold comparable to the telescope resolution at that band.
These results demonstrate the readiness of this pipeline for incorporation into upcoming Simons Observatory data analyses.

\end{abstract}

\section{Introduction}
\label{sec:intro}
Highly energetic and explosive astrophysical phenomena, including super-novae (SNe), binary star mergers, tidal disruption events (TDEs), gamma ray bursts (GRBs), active Galactic nuclei (AGN), and pulsars, feature relativistic jets that drive strong shock fronts \citep{Phinney1991, Weiler2002, Totani2002, Chandra2012, Gair2014, Fraser2020, Alexander2020, Somalwar2022}.
These shocks accelerate charged particles in the surrounding plasma to relativistic speeds, producing synchrotron radiation. 

Stellar flares also generate synchrotron emission through magnetic reconnection events, accelerating particles in strong magnetic fields near the stellar surface \citep{Hawley2016,kowalski2024}.
The resulting synchrotron emission from these phenomena presents as transient point sources across cosmic microwave background (CMB) frequency bands (30--300\,GHz), with spectral energy distributions that are particularly bright at low frequencies relative to the peak CMB signal.

Ground-based CMB experiments have been observing the early Universe for over 60 years, and are presently scanning large fractions of the sky at millimeter wavelengths with arcminute resolution.
As these experiments have expanded beyond traditional cosmological observations, modern CMB observatories such as the Atacama Cosmology Telescope \citep[ACT; ][]{Aiola2020} and the South Pole Telescope \citep[SPT; ][]{Carlstrom2011} have demonstrated the potential for millimeter-wave time-domain science \citep{Eftekhari2022}.
The Simons Observatory Large Aperture Telescope (SO LAT) will significantly advance this field by building upon the strengths of both ACT and SPT through wide sky coverage, high instantaneous sensitivity, multi-frequency observations and a large field of view \citep{Ade2019}.

Although radio detections and follow-up observations of transients have historically been focused at centimeter to meter wavelengths, synchrotron emission does extend to millimeter scales. 
The millimeter-wave transient sky, however, is largely unexplored.
Only recently have fast or early ($\lesssim\mathcal{O}[\mathrm{days}]$-old) transients detected in optical surveys seen follow-up observations at millimeter-wavelengths via sensitive millimeter interferometers such as the Atacama Large Millimeter Array \citep{Laskar2013, Laskar2016, Laskar2019, Ho2019, Maeda2021, Andreoni2022, Berger2023}.
These, along with the first direct map-based detections of transient events from SPT and ACT, have allowed for early estimates of millimeter-wave flux density limits, and provided constraints on the properties of the associated sources, demonstrating the need to actively survey this frequency band \citep{Whitehorn2016, Guns2021, Naess2021, Li2023, Tandoi2024, Hervias2024, Biermann2024}.

By 2028, the SO LAT \citep{Xu2021} will be capable of routinely and efficiently discovering transients in the millimeter-wave sky, generating daily maps for transient detection similar to those used in \cite{Biermann2024}, and producing light curves for each identified source to be disseminated to the community within $\sim30$\,hours \citep{Abitbol2025}.
Millimeter-wave transient surveys from the SO LAT will allow us to probe the physical properties of GRB outflows through reverse shock emission, independently measure the event rate of TDEs hosting relativistic outflows, probe the physics of shock acceleration, and constrain the intrinsic rates of millimeter-wave transients \citep{Piran1999, Laskar2013, Laskar2016, Laskar2019}.

Time-ordered data from CMB surveys are typically integrated into maps, averaging signals over timescales of a few minutes \citep{Dunner2013, Dutcher2018, Aiola2020, Naess2025}.
To probe higher time resolution, and access rapidly evolving astrophysical transients, we need to directly explore the time-domain.
Stellar flares, for example, as observed from SPT and ACT map-based methods, are predicted to be the most common class of transients detected with the SO LAT \citep{Abitbol2025}.
Although these flares often take a few hours to rise to and fall from their peak flare amplitude, some last only a few minutes \citep{Biermann2024}.
Similarly, reverse shock components of the afterglow emission from long GRBs are expected to peak in the millimeter on scales ranging from hours \citep{Laskar2019} down to minutes \citep{Eftekhari2022}.
Even slowly rotating pulsars observed at $\sim100$\,GHz will peak over periods of 0.1 to 10\,s \citep{Torne2021}.

Searches for fast (seconds-minutes), millimeter-wave transients can probe the physics of both stellar flares and GRBs. 
For GRBs, rapid millimeter follow-up can reveal short-lived reverse-shock emission while it is still bright, and potentially dominant over the forward shock, providing a direct diagnostic of the ejecta properties, including its magnetization and initial Lorentz factor \citep{Zhang2005,Laskar2013,Perley2014}. 
Similarly, rapid detections of stellar flares can capture the impulsive phase of magnetic reconnection, providing constraints on flare energetics, magnetic-field environments, and particle acceleration, while also informing the high-energy radiation and space-weather conditions experienced by planets around active stars \citep{MacGregor2018, Howard2022}.

In this paper, we present a method for blindly detecting transient events in SO LAT time-ordered data.
By directly exploring the time-ordered data, our approach avoids the need to construct maps, enabling sensitivity to transients evolving on timescales too short to be resolved in map space.
We also present this pipeline in the context of the Simons Observatory (SO) time-domain alert system, for which we will be able to release public astrophysical alerts generated through timestream or map-based blind searches.
This allows experiments such as SO to play an important role in the larger astrophysical community, filling in one of the less-explored ranges of transient science across the electromagnetic spectrum. 

In Section~\ref{sec:data} we detail the SO LAT and its optics. 
We also describe our simulated observations and the methodology of injecting flare-like light curves.
In Section~\ref{sec:processing} we describe the steps to prepare time-ordered data, and discuss the detection of astrophysical sources in Section~\ref{sec:detecting}.
The pipeline is then applied to the simulated observations in Section~\ref{sec:forecasts} to estimate detection thresholds and completeness.
We conclude in Section~\ref{sec:conclusion}.

\begin{table}
    \centering
    \caption{SO LAT receiver specifications}
    \begin{tabular}{c|c|c|c|c}
    \toprule
    \toprule
        Frequency & FWHM & Frequency & Detector & Optics \\
        $\mathrm{[GHz]}$ & $\mathrm{[arcmin]}$ & Bands & Count & Tubes \\
    \midrule
        27 & 7.4 & LF & 222 & 1\\
        39 & 5.1 & & 222 & \\
    \midrule
        93 & 2.2 & MF & 10,320 & 4\\
        145 & 1.4 & & 10,320 & \\
    \midrule
        225 & 1.0 & UHF & 5,160 & 2\\
        280 & 0.9 & & 5,160 & \\
    \bottomrule
    \end{tabular}
    
    \vspace{0.5em}
    {\small Expected instrumental properties for the initial deployment of the SO LAT receiver, as detailed by~\cite{Zhu2021}. The listed observing frequencies and beam FWHM are approximate.}
    \label{tab:sens}
\end{table}

\section{Simulating Simons Observatory Data}
\label{sec:data}

We begin this section with an overview of the SO LAT receiver and survey strategy. We then introduce the simulations of SO LAT observations and the flare-like light curves that are injected into the simulated time-ordered data to characterize the transient detection pipeline.

\subsection{The Simons Observatory Large Aperture Telescope Receiver}
\label{subsec:SO}

The 6\,m aperture SO LAT will survey $\sim61$\,\% of the sky with over 60,000 dichroic transition edge sensor detectors \citep{SOWhitepaper,Xu2021,Zhu2021,Abitbol2025}.
The initial deployment of 7 optics tubes will allow near-simultaneous observations in six frequency bands: one low frequency optics tube (LF -- 27 \& 39\,GHz), four medium frequency optics tubes (MF -- 93 \& 145\,GHz) and two ultra-high frequency optics tubes (UHF -- 225 \& 280\,GHz), with angular resolution and specific detector counts shown in Table~\ref{tab:sens}.

Each optics tube contains three dual-polarization, dual-frequency detector wafers, giving a total of 3 LF, 12 MF and 6 UHF wafers for the initial configuration.
The individual MF and UHF wafers will host 430 pixels per frequency band, and the LF wafers consist of 37 pixels per channel (or 74 detectors) due to their larger physical size.
Pixels in this context correspond to the dual-polarization combination of detectors, so each pixel consists of two detectors, one for each of two orthogonal linear polarizations. 

\begin{figure}
    \centering
    \includegraphics[width=\linewidth]{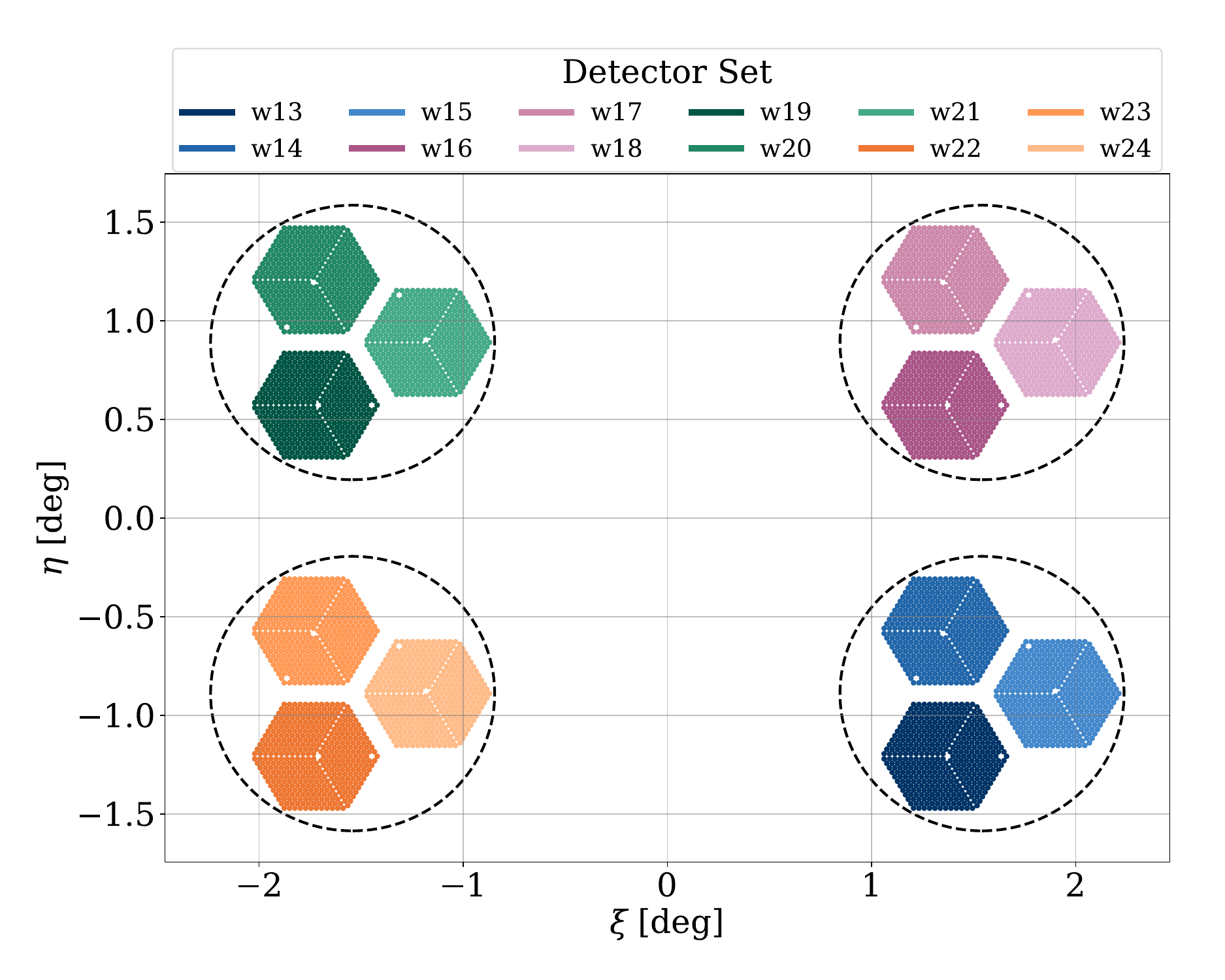}
    \caption{
    An example layout of the focal plane coordinates ($\xi,\eta$) of pixels in the SO LAT. 
    This configuration shows four simulated MF optics tubes (dashed circles), each consisting of three hexagonal detector wafers, colored by wafer ID.
    Note the wafer IDs here are arbitrarily set by the simulations.
    Each colored point within a hexagon denotes a unique pixel (for each frequency a pixel includes two detectors: one per polarization direction).
    We can use these layouts to relate the focal plane coordinates of each pixel to approximate where they are pointed on the sky.
    }
    \label{fig:full_focalplane}
\end{figure}

In Figure~\ref{fig:full_focalplane} we show a representation of four simulated MF optics tubes, each hosting three detector wafers.
The coordinates $\xi$ and $\eta$ represent angular offsets from the boresight of the telescope in a flat projection.
The physical position of each detector in the focal plane directly determines the angular direction it observes on the sky, establishing a mapping between the focal plane position and sky angle.
This allows us to consider Figure~\ref{fig:full_focalplane} as simultaneously representing the detector layout in the focal plane and the telescope's instantaneous response to the sky.
Each wafer consists of 860 pixels, able to simultaneously detect at two frequencies (93 and 145\,GHz in the MF example shown), or 430 dual-polarization pixels per frequency.
Note that the IDs attributed to each wafer are an arbitrary labeling set by the simulations. 

Each optics tube has a field of view of $\sim1.3^\circ$ in diameter, so in this example configuration we have a total field of view of $\sim16$ square degrees.
The total field of view of the fully populated SO LAT receiver will be approximately $7.8^\circ$ in diameter \citep{Ade2019}.
In this full configuration, the sensitivity to point sources will have a baseline (goal) level of 9.5\,mJy (6.9\,mJy) within a single observation at 93\,GHz \citep{Abitbol2025}. 
Performing constant elevation scans and mapping the full focal plane leads to a map integration-time of 30 minutes on a source. 
For a single wafer, $\sim0.5^\circ$ in diameter, these sensitivities then scale to a baseline (goal) of 46.5\,mJy (33.8\,mJy), with map integration-time of $\sim3$ minutes.
While the optics tubes field of view and specifics of Table~\ref{tab:sens} are the true measured values \citep{Zhu2021}, the configuration of the optics tubes and wafers shown in Figure~\ref{fig:full_focalplane} are assumed from our simulations and will not be exactly the same as the true focal plane of the SO LAT after commissioning.

\subsection{Survey Strategy}
\label{subsec:survey_strat}

The SO LAT will be operated with a high-cadence, wide-area survey strategy.
While the final scan strategy is still being optimized, ground-based observatories, such as ACT, typically use constant elevation scans, sweeping back and forth in azimuth.
These provide uniform coverage within each scan, helping to mitigate systematic effects such as atmospheric noise \citep{Stevens2018}.

Assuming a constant elevation scan strategy, the telescope is expected to scan at an azimuthal rate of $1^\circ$ per second (projected on the sky), balancing rapid sky coverage with sufficient integration time per sky pixel.
At this rate, the telescope would cover approximately 16,000 square degrees of sky (roughly 40\%) per day \citep{Bhandarkar2025, Abitbol2025}, enabling a high-cadence survey crucial for detecting astrophysical transients.

Observations will be recorded in $\sim1$\,hour segments of time-ordered data per detector in each wafer.
The observations are sampled at $220$\,Hz in the MF bands and $340$\,Hz in the UHF bands, corresponding to time sample spacings on the sky of $16.4^{\prime\prime}$ and $10.6^{\prime\prime}$, respectively.
For comparison, the average pixel spacing is $1.57^\prime$ in both bands.
Note that the final scan pattern and scan speed of the survey are under evaluation during commissioning and initial science observations, and may change from the simulated configuration we have used here.

A point source traces an approximately horizontal path across our simulated focal plane in Figure~\ref{fig:full_focalplane} because the telescope scans at a fixed elevation.
Without loss of generality, assume the telescope is slewing from left to right.
A source that enters the focal plane at $\eta=-1$ will first be observed by the bottom left optics tube, passing through wafer 22 into wafer 24, crossing through the center of the focal plane before reaching the bottom right optics tube and being observed by a row of pixels in wafer 13 and wafer 15, finally leaving the focal plane on the right hand side.
At the above scan rate, this entire focal plane crossing would take approximately 5\,s, where each individual wafer would see the source for between 0.3--0.5\,s.
We will refer back to these values in our discussion of transient detection cuts.

It is also worth noting that at higher frequencies of light, ground-based CMB observations become increasingly affected by atmospheric emission and absorption, particularly beyond 150\,GHz where strong water vapor lines dominate \citep{Errard2015}.
This increased opacity both raises the optical loading on detectors (amplifying photon noise) and introduces stronger, more correlated atmospheric fluctuations in time-ordered data \citep{morris2022}.
These fluctuations are modulated by the scanning strategy (e.g., scan speed and direction relative to any present wind) and atmospheric structure (e.g., turbulence scales).
During initial timestream calibration, some atmospheric emission will be removed through simple high-pass filtering; however we still expect sensitivity in time-ordered data to decrease at higher frequencies, particularly in the UHF bands.
Considering also the significantly larger beam sizes and reduced detector counts in the LF bands, we expect the best sensitivity to sources will be at 93 and 145\,GHz.

\subsection{Simulated Observations and Events}
\label{subsec:sims}

\begin{figure}
    \centering
    \includegraphics[width=\linewidth]{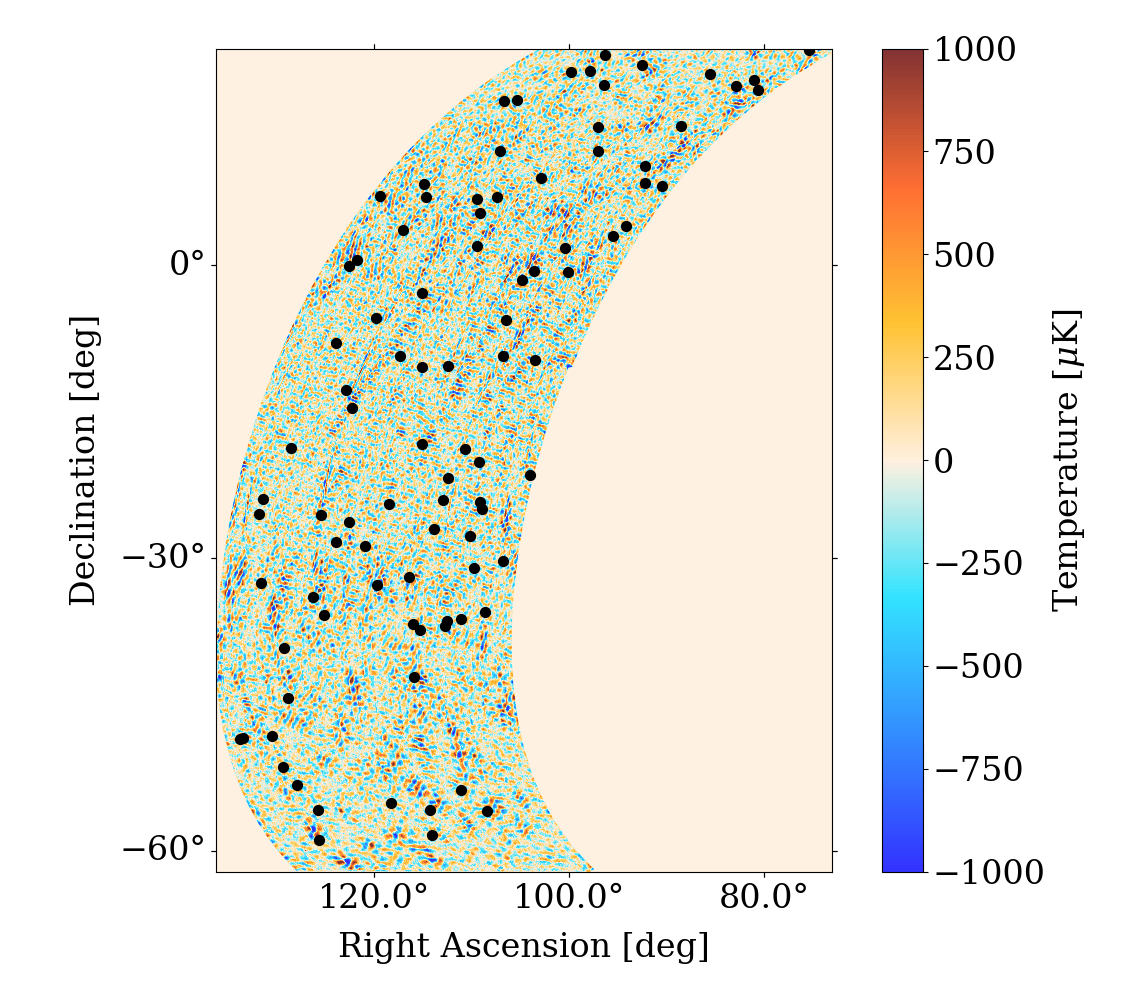}
    \caption{
    A map from a single wafer for the two hours of simulated observations, with the locations of injected, simulated transient sources plotted as black circles. 
    The simulation includes CMB, white noise, and atmospheric emission.
    }
    \label{fig:radec_sim_example}
\end{figure}

We test the transient detection pipeline using two hours of simulated time-ordered data from SO LAT observations in a nominal survey configuration. 
The simulation is based on the nominal, wide 3-elevation observing strategy. 
Realistic boresight motion is coupled with a synthetic CMB signal, atmospheric emission, azimuth-synchronous signal and elevation-modulated instrumental noise.  
Gain miscalibration was added by drawing Gaussian gain errors with $1\,\%$ standard deviation. 
The simulated time-ordered data were generated with this configuration using the Simons Observatory: Time Ordered Data Library (\texttt{SOTODLib}\footnote{\url{https://github.com/simonsobs/sotodlib}}), and Time-Ordered Astrophysics Scalable Tools (\texttt{TOAST}\footnote{\url{https://github.com/hpc4cmb/toast}}).

To simulate astrophysical transient events, we generate toy-model light curves inspired by those observed at millimeter-wavelengths \citep{Guns2021,Tandoi2024,Biermann2024}.
These flares are commonly fit with Gaussian profiles \citep{MacGregor2018} or exponential decay curves \citep{Salter2010}, so we opt to use a simple power-law model for the transient rise and decay, 

\begin{equation}\label{eq:flaresim}
    S_{f}(t) =
        \bigg\{
    \begin{array}{lr}
        S_{p}\left(\frac{t+t_{r}-t_{p}}{t_{r}}\right)^{\eta_{r}}, & t \le t_{p}\\
        S_{p}\left(\frac{t_{p} + t_{d} - t}{t_{d}}\right)^{\eta_{d}}, & t > t_{p}
    \end{array},
\end{equation}

\noindent
where $S_f(t)$ is the flux density at time $t$, and $\eta_r$, $\eta_d$ are the power-law indices for the rise and decay, respectively.
The parameters $t_p$, $t_r$, and $t_d$ correspond to the time of peak flux density, the rise duration, and the decay duration, with $S_p$ being the peak flux density at $t_p$.

The choices of $\eta_r$ and $\eta_d$ affect the sharpness with which the flare rises to or decays from the peak. 
A larger power will result in a faster rise/decay and longer tail, while setting these to one would represent a linear rise/fall. 
For our later tests, the main goal is the detection of the peak itself, so we set $\eta_r=\eta_d=4$ to represent moderate but balanced rise and decay shapes. 
We also set symmetric rise and decay times for the same reason and to ensure consistency without blowing out the parameter space when forecasting detection rates. 
In practice, flaring events do not have symmetric rise and decay rates or shapes, for example, stellar flares often have longer and flatter decays than rises. 
While this can affect how much of the flare light curve can be observed and hence modeled, neither will directly impact observation of the peak itself.

\begin{figure}
    \centering
    \includegraphics[width=\linewidth]{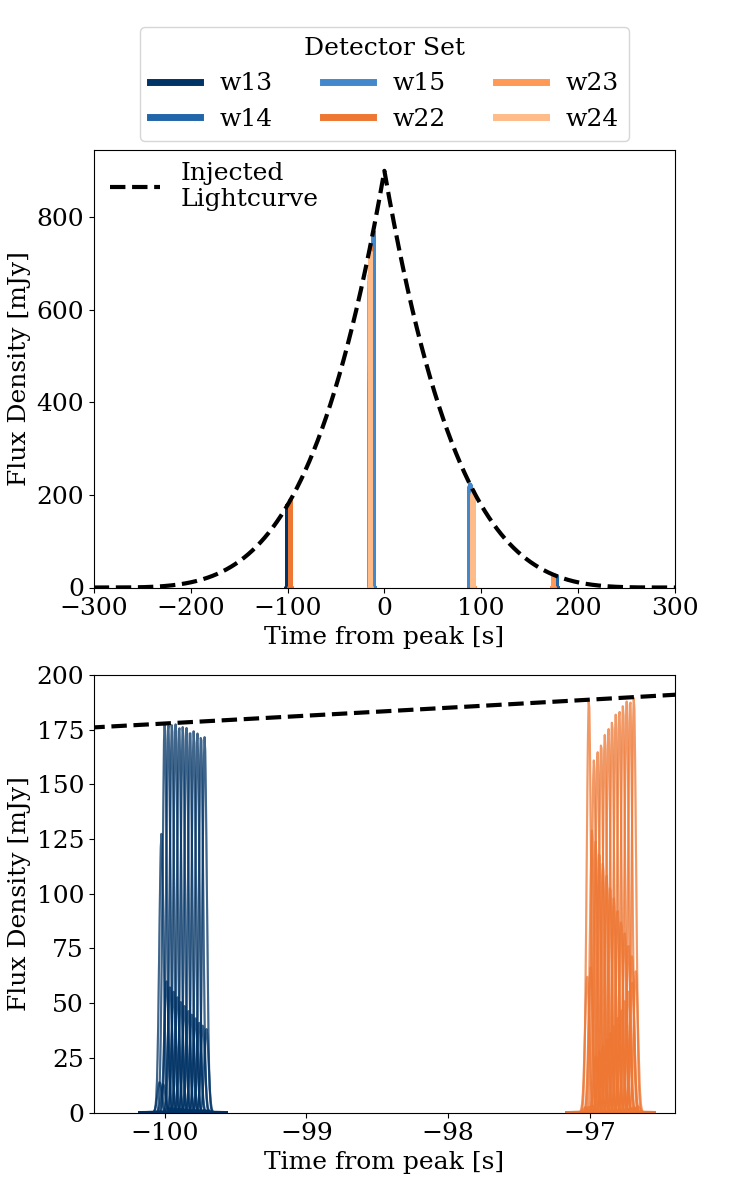}
    \caption{
    Light curve of a simulated flare (black) and the signal response (colored by wafer -- matching Figure~\ref{fig:full_focalplane}). The injected signal response is shown as time-ordered data from all detectors, in all simulated MF wafers. 
    We show the response from all wafers that observe the source over its entire duration (top), and a zoom in of the first time the source is observed in an individual optics tube (bottom).
    The grouping of individual wafer detections separated by $\sim80$--100\,s corresponds to the source entering, passing across, and exiting the focal plane from either left or right directional scanning.
    This is clear in the zoom-in, where we have two wafers from neighboring optics tubes separated by $\sim2.5$\,s, corresponding to the distance of $\sim2.5^\circ$ between wafers on the focal plane.
    }
    \label{fig:flux_sim_example}
\end{figure}

We inject flares into the simulated observations at a position and time randomly sampled from the constant velocity portion of scans. 
The constant velocity portion includes 97\% of the time-ordered data in these simulations, and is expected to cover 97-99\% of the time-ordered data in actual observations depending on the final scan strategy. 
We then smooth by the instrument's beam response, $\theta_\mathrm{FWHM}$, at each frequency, ensuring the injected flares represent varying point sources.
Figure~\ref{fig:radec_sim_example} shows a sample of 100 events (black circles), spread across a map made from the simulated observations.

\subsection{Temporal Signal Characteristics}
\label{subsec:time_characteristics}

As the telescope scans across the sky, transient sources periodically enter and exit the focal plane.
Figure~\ref{fig:flux_sim_example} demonstrates this process, where the black dashed line shows the injected light curve and the colored lines represent detector coverage across all wafers (colored by simulated wafer ID, matching Figure~\ref{fig:full_focalplane}).
In this example, the source is clearly detected on three distinct occasions over the duration of the event (top panel) -- we have three scans across the full focal plane from Figure~\ref{fig:full_focalplane}. 

The timing between detector passes reflects the relationship between the telescope scan pattern and the source position on the sky, where wide gaps ($\sim80-100$\,s for the flare in Figure~\ref{fig:flux_sim_example}) occur when the source image leaves the focal plane (exiting from the left or right side of Figure~\ref{fig:full_focalplane}) until the instrument completes a full scan, turns around and returns from the other direction.
The relative asymmetry in intervals is due to the source's position relative to the turnaround points and scan center.
A non-varying source that lies at the scan center will show symmetric intervals between observations, while one located away from the center will have asymmetric patterns like the 80\,s and 100\,s gaps seen in Figure~\ref{fig:flux_sim_example}.

As described in Sections~\ref{subsec:SO} \&~\ref{subsec:survey_strat}, it will take 0.3--0.65\,s for a source to cross a single wafer at an azimuthal scan rate of $1\,\mathrm{deg}\,\mathrm{s}^{-1}$ (the angular speed as projected on the sky).
With gaps being present between each optics tube we find it takes 2--3\,s for a source to pass between adjacent optics tubes. 
This can be seen in the bottom panel of Figure~\ref{fig:flux_sim_example} where, across the $\sim4$\,s window, the source is observed by two wafers with a $2.5$\,s gap between.
If we reference back to Figure~\ref{fig:full_focalplane}, this represents the source passing from right to left, from wafer 13 in the bottom right optics tube to the next adjacent set of detectors, wafer 22 in the bottom left optics tube.

We can also observe the individual detector responses plotted atop one another in the bottom panel of Figure~\ref{fig:flux_sim_example}.
The scaling of each detector's Gaussian beam response relates to the spatial offset from the source's path across the focal plane.
As the source isn't likely to travel perfectly down the center of a row of pixels, some detectors will have larger offsets and thus measure a reduced flux density.

In the example of Figure~\ref{fig:flux_sim_example}, the true peak of the injected light curve is never sampled by any of the detector wafers.
For short-duration events, the scan may pass over the source before the flare begins, and the emission can rise and decay entirely between successive passes.
Consequently, even intrinsically bright events can be missed, not due to sensitivity limits, but because of the interplay between the scan cadence and the source's variability timescale.
This introduces an unavoidable element of chance -- our ability to detect these events will always have some reliance on looking at the right place at the right time.

These sampling effects can also introduce a variation in characterizing transients dependent on the scan position. 
For events with durations comparable to or shorter than the time between successive focal-plane crossings, depending on where a source is observed within the directional scan pattern, we might see closely packed samples, widely separated samples, or only a single effective pass. 
This could lead to variations in recovered peak flux, localization, light-curve sensitivity, and potentially detection completeness. 
While we do inject simulated flaring events to cover varying scan positions, we do not estimate this effect quantitatively and leave it for future work.

\section{Processing Time-Ordered Data}
\label{sec:processing}

In this section, we will describe the filtering done to increase the detection signal to noise (S/N) of transient candidates.

\subsection{Pre-processing}
\label{subsec:preprocess}

The standard conversion from time-ordered data packages into CMB maps involves careful data quality cuts and calibration; processes standardized by past and current experiments \citep{Dunner2013,Dutcher2018,Aiola2020,Naess2025}.
These remove significant spurious signals such as glitches (cosmic rays, electromagnetic interference, electronic readout, etc.), cut highly contaminated detectors, and calibrate the data units from raw detector output to relative CMB temperature fluctuations.

It is important to note that rapidly varying astrophysical sources could be bright enough to be flagged as a glitch given they appear as an intense signal burst.
Glitches are typically flagged when a set of samples has S/N$~\ge10$.
A snippet around the flagged samples will then be removed from the time-ordered data.
Improving glitch cutting procedures is an active area of development, particularly in the analysis and classification of glitches that may originate from point sources and transient events to avoid these being lost as systematics.
\cite{Nerval2025} for example, use a machine learning approach to classify glitches and separate astrophysical or instrumental phenomena.

In the search for rapid and dynamic astrophysical transients, satellite contamination is also becoming a significant challenge.
Satellites can mimic short-duration events in time-resolved survey data, appearing as either glitch-like spikes in time-ordered data, or as bright streaks in high-cadence maps.
In CMB channels, these high S/N signatures are primarily due to thermal emission from the satellites.
This issue is expected to be particularly prevalent for experiments located beneath dense equatorial orbits, such as SO, given the rapid growth in satellite constellations. 

While the true impact of today's satellite population on millimeter-wave time-domain studies is unknown, early detections have already begun to show difficulties arising in our potential abilities to mask or cross-check satellites due to discrepancies in their ephemerides \citep{Foster2024}. 
Modeling, tracking and understanding how satellites and satellite constellations will affect astronomy across the electromagnetic spectrum is an ongoing issue \citep{Peel2025,Dadighat2024}.
Although we do not consider satellites in the simulations of our work, this pipeline will be a useful tool in estimating the significance of contamination by satellites, and how important it will be to consider these in future CMB data-processing. 
The source-geometry cuts described in Section~\ref{subsec:cluster_cuts} may reject satellites with apparent motion inconsistent with the expected path of a point source across the focal plane.
However, the separation between astrophysical transients and satellites is not tested in this work. 

\subsection{Filtering}
\label{subsec:filters}

\begin{figure}
    \centering
    \includegraphics[width=\linewidth]{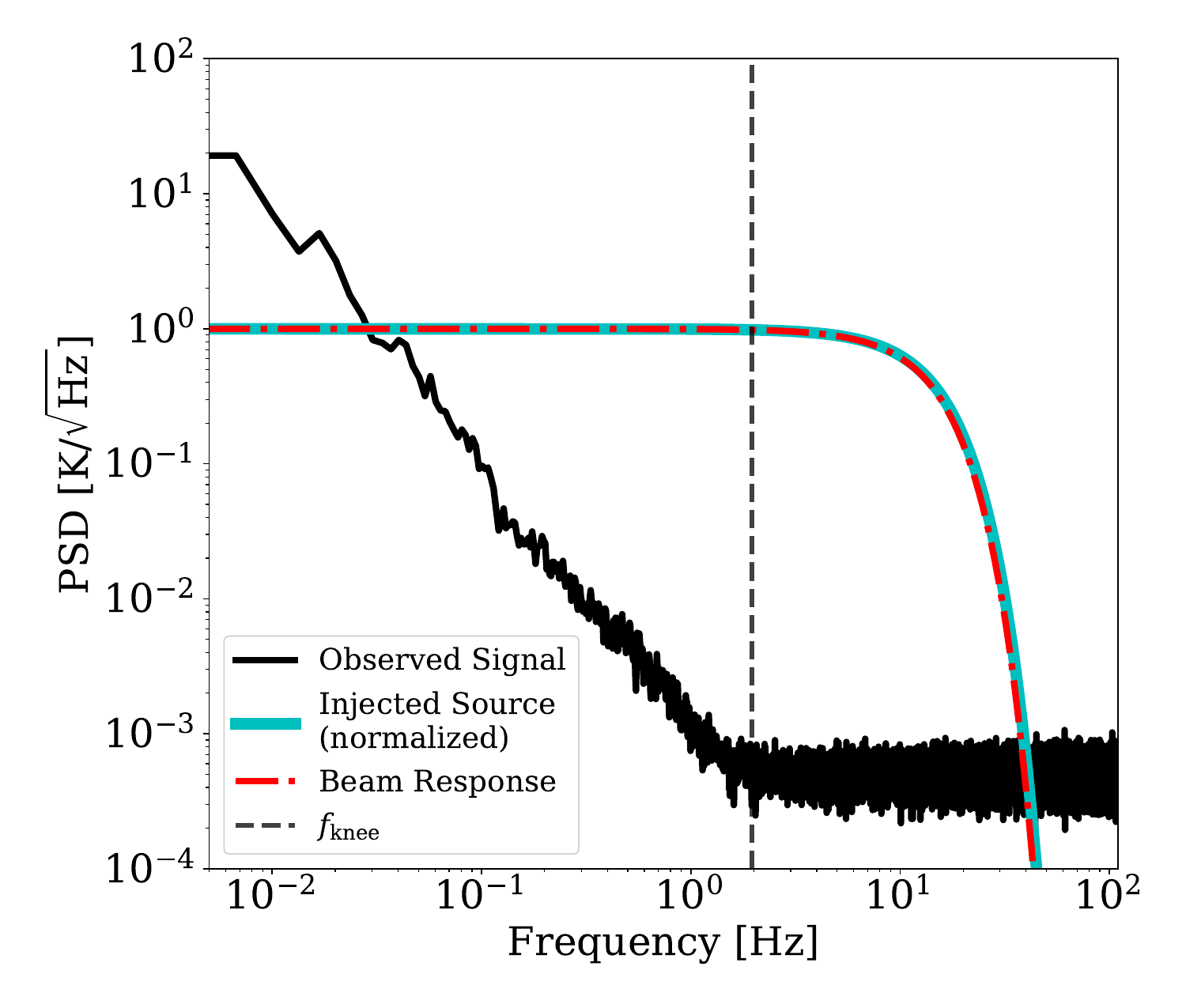}
    \caption{
    An example of the frequency response of a single detector (black) at 93\,GHz with strong red noise, primarily from atmospheric contamination, at low frequencies and white noise across all scales. The normalized beam function has been overlaid (dashed red) with the (almost identical) clean signal response to a single point source (cyan), also normalized to easily compare the signal structure. The transition from red noise to mostly white is shown as the dashed black line, denoting $f_\mathrm{knee}=1.23$\,Hz.
    }
    \label{fig:fft_loglog}
\end{figure}

Time-ordered data, after pre-processing, remain contaminated by noise across all frequencies, which limits our ability to detect signals at the scale of point sources.
A large source of residual noise is atmospheric emission, which introduces significant red noise, following a steep power-law spectrum and projecting onto large angular scales \citep{Amblard2004, morris2022}.
In addition, instrumental white noise contributes a flat power spectral density (PSD) across all frequencies, obscuring faint signals, particularly at small angular scales.  

These noise contributions can be seen in Figure~\ref{fig:fft_loglog}, which shows the PSD of a single 93\,GHz observation in black.
The steep atmospheric contamination dominates up to $\sim2$\,Hz (explicitly, this depends on the atmospheric realization where, during observations with good weather, it will be lower).
We also show the PSD of a simulated event that would be injected into this signal in cyan, and the instrument's beam response as a dashed red line.
The PSD of a point source is determined by convolution with the beam shape, as is evident by the overlap of these lines. 

To suppress the remaining noise contaminants and enhance the S/N of sources throughout the time-ordered data, we apply a matched filter. 
Matched filtering is a standard technique in astrophysical signal processing and is widely used in CMB experiments for detecting point sources and galaxy clusters \citep[eg., ][]{hilton2021}.
This technique detects the presence of a pre-defined template, $T$, in a noisy signal, $h$, by correlating the two in a way that maximizes the S/N with respect to $T$ \citep{Turin1976, Vio2004, Hernandez2008, Zubeldia2021}.

Since we are searching for point source transients, we use a Gaussian matching the beam profile as our template.
While the true beam profile will not be perfectly Gaussian, this is a good simplification without requiring complex side-lobes or jitter that we can safely assume to be negligible in the context of our filter.
The filter is then defined as,

\begin{equation}\label{eq:matched_filter}
    (T|h)(t)=\mathbb{R}\left(\int^\infty_0df\frac{\tilde{T}^*(f)\tilde{h}(f)}{S_n(f)}\mathrm{e}^{2i\pi ft}\right),
\end{equation}

\noindent
where $\tilde{T}(f)$ and $\tilde{h}(f)$ are the Fourier transforms of the template and signal at Fourier frequency, $f$, respectively.
The asterisk denotes complex conjugation, $S_n(f)$ is the noise PSD estimated from the signal, and $t$ is the evaluation time.
The real part is taken to obtain the physical (real-valued) filtered signal, $\mathbb{R}$, where the exponential appears in the inverse Fourier transform to bring the filtered signal back from frequency space to the time-domain.

Since time-ordered data contain noise across all scales, a generic matched filter can inadvertently amplify poorly constrained modes. 
The template's beam response, as seen in Figure~\ref{fig:fft_loglog}, acts as a low-pass cutoff, suppressing high frequency noise, for example at $f\gtrsim40$\,Hz in the 93\,GHz band.
The noise PSD, $S_n(f)$, works to generically suppress frequencies where the noise is large, i.e., low frequencies with high atmospheric contamination; however, this relies on a well-estimated PSD, which is not always possible in time-ordered data.

If the suppression from the estimated noise PSD is not sufficient or stable, it can lead to numerical instabilities where $S_n(f)$ is very low or poorly constrained, or leakage from non-signal modes such as long-timescale drifts at low frequencies or high frequency digitization noise. 
To avoid this, we apply a high-pass Butterworth filter at the atmospheric knee frequency ($f_\mathrm{knee}=1.23\pm0.54$\,Hz at 93\,GHz), where the red noise transitions to white.
This forces low frequencies to zero where poorly constrained power could contaminate the filtered signal.
The band-limited matched filter then acts as a scale-selective, noise-weighted convolution that suppresses both large- and small-scale noise while maximizing S/N for point source detection.

It is important to note that Fourier filters tend to keep the average of a timestream constant which can cause spikes at the beginning and end of the signal and a subsequent reduction of amplitude.
If these filter spikes appear in any of our timestreams, they'll look clearly distinct from Gaussian astrophysical point sources and will not survive the cuts described in Section~\ref{subsec:cluster_cuts}.
We also ran a Monte Carlo simulation of 5000 flares to estimate the suppression in the peak signal from the Fourier filters, showing a recovery of $85.01\pm0.02$\% of the input flux.
However, this $\sim15$\% suppression is largely offset by an overestimation from our final flux fits described in Section~\ref{subsec:rates}.

\section{Detecting Source Transits in Time-Ordered Data}
\label{sec:detecting}

In this section, we describe the steps to identify and extract localized candidate events from the match-filtered timestream, based on the detection of high S/N samples during individual telescope scans.
Given that the S/N of the match-filtered data quantifies the degree of correlation between the signal and the beam response over time-ordered data, these events represent single scan transits of point sources across the focal plane, independent of any assumptions about source variability.

Sources may produce multiple such detections, but here we focus on characterizing the detection of each individual crossing to achieve optimal time-resolution, allowing us to later characterize the variability in detail.

\subsection{Identifying Candidate Sources}
\label{subsec:cluster}

We begin by selecting timestream samples with matched-filtered S/N$\,\ge3$ from all detectors in a given wafer.
These samples are then grouped based on their corresponding sky coordinates, with a typical spatial scale set by the beam at a given frequency.

To identify where the telescope is looking at each time sample, we use a pointing model that maps the detector samples to celestial coordinates. 
This is represented by a pointing matrix, which accounts for the telescope's time-dependent boresight trajectory, detector offsets within the focal plane, and the observing time.
For each detector and time sample, this mapping yields the right ascension (RA), declination (Dec), and the detector orientation on the sky.

As we noted above, to maximize our time resolution we want to detect a single-scan transit of a source.
To isolate single scans we segment the time-ordered data at the telescope turnaround points and treat each scan (typically $\sim80$--$100$\,s in duration) independently.
For slower flares, like the example given in Figure~\ref{fig:flux_sim_example}, this means we can observe the source multiple times during the one set of $\sim$hour long data.
Faster flares may be limited to a single scan but may still pass over multiple detector wafers as the source crosses the full focal plane.

Within a scan, we extract the RA and Dec for each high S/N sample and apply the \texttt{scikit-learn}\footnote{\texttt{scikit-learn}: \url{https://scikit-learn.org/stable/index.html}
\citep{Pedregosa2011}} density-based spatial-clustering-of-applications-with-noise algorithm \citep[DBSCAN; ][]{Ester1996}.
This takes a set of samples, groups together points that are closely packed (many neighbors), and marks outliers in low-density regions, clustering samples based on a minimum spatial scale.

The characteristic spatial scale of a point-source is known from the beam size. 
We can apply well-characterized beams to DBSCAN to act as a `neighborhood radius`, grouping connected samples within this scale while simultaneously identifying isolated samples as noise.
We define this spatial clustering scale as the beam width, $\sigma_\mathrm{FWHM} =\theta_\mathrm{FWHM}/(2\sqrt{2\ln{2}})$, to reflect the angular scale over which signal from a point source is expected to be concentrated.
This is frequency-dependent due to the band-specific beam sizes (see Table~\ref{tab:sens}).

To ensure reliable clustering, we retain only groups containing more than 10 high-S/N samples from across a wafer. 
DBSCAN clusters candidate sources across all detectors at once, so for the example case of a minimum of 10 samples, these could all come from a single detector, or there could be a single sample from 10 unique detectors.  
This threshold balances the need for sufficient information to distinguish true signals from noise. 
Reducing the sample requirement increases the number of detected groups but risks fragmenting real sources and increasing contamination by noise fluctuations, while raising the sample requirement reduces sensitivity to events that produce only brief or sparsely sampled responses. 
We discard the low-density outliers and the remaining groups of samples are retained as candidate source events for further analysis.

\subsection{Candidate Source Cuts}
\label{subsec:cluster_cuts}

\begin{figure}
    \centering
    \includegraphics[width=\linewidth]{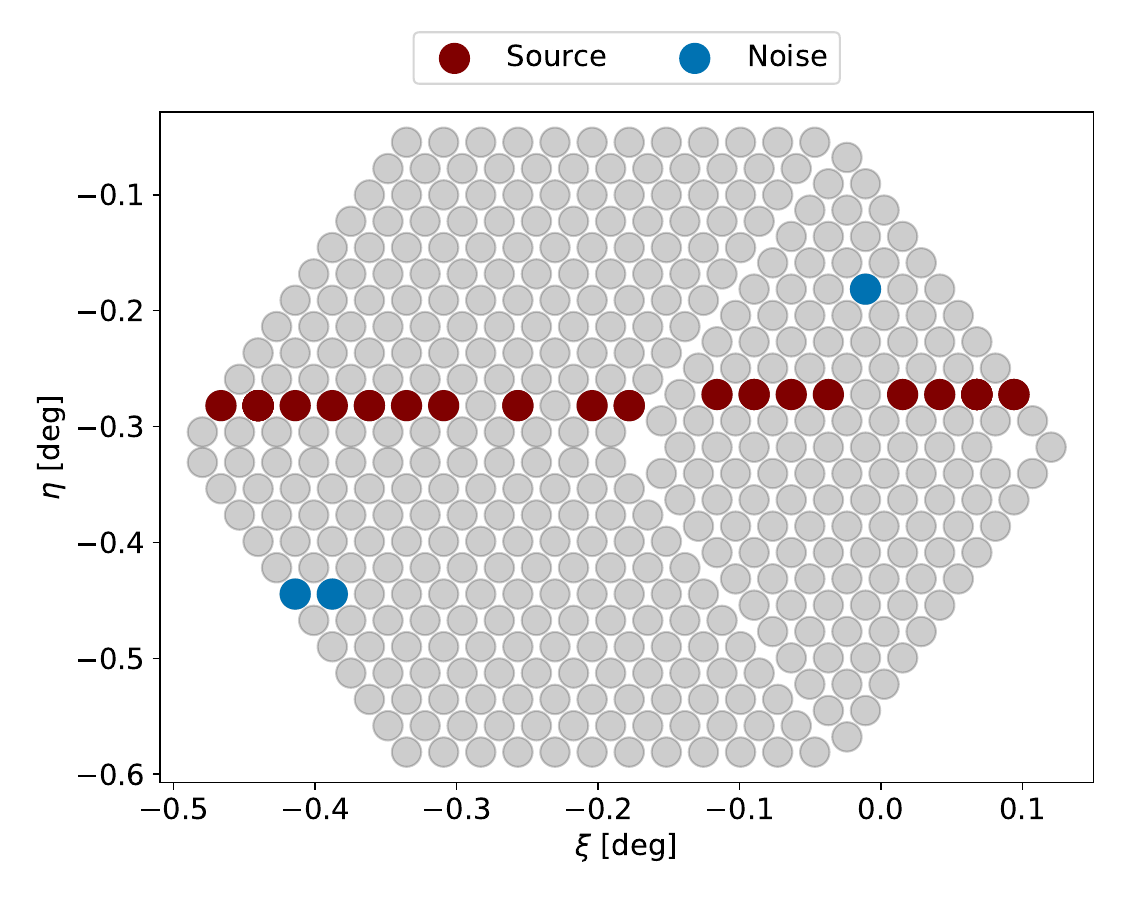}
    \caption{
    Pixel distribution across the focal plane of a single MF wafer (gray) with flagged pixels representing a point source (red) and two examples of noise (blue). The source falling along a line of constant elevation is due to the telescope scanning at constant elevation.
    }
    \label{fig:focal_plane}
\end{figure}

To distinguish true scans of a source from noise or other systematics, we apply a series of selection cuts.
These include thresholds on the number of detectors the event was flagged across, geometric checks to identify readout glitches, and constraints on source amplitude and localization.

As the telescope scans at fixed elevation, a source should appear as a linear feature at roughly constant elevation across the focal plane.
Figure~\ref{fig:focal_plane} illustrates the pixel layout for one of our simulated MF wafers, where a point source (red) forms a coherent track across the length of the detector array, and spurious noise (blue) would appear spatially inconsistent.

The DBSCAN parameters constrain the total number of samples in a candidate source, but do not constrain the number of unique detectors that contribute to those samples. For example, a cluster containing 20 samples could pass the DBSCAN requirement even if all samples originated from only two detectors.
We therefore require that each candidate event consist of at least five individual detectors and assess the spatial geometry of the detectors on the array to ensure the response is `source-like'.
This means the flagged detectors form a straight line with a large aspect ratio (the ratio of the length of the line of detectors to the height should be large, as can be seen by the colored pixels in Figure~\ref{fig:focal_plane}).
A source candidate that fails this will be dropped as noise.  

As discussed, due to the dual-frequency, dual-polarization design of the transition-edge sensor detector wafers \citep{Stevens2020}, each point in Figure~\ref{fig:focal_plane} represents two pixels, or four detectors; two with orthogonal polarization sensitivity per frequency (e.g., two detectors at 93\,GHz and two at 145\,GHz).
Considering this, a highly polarized source will be restricted to fewer detectors per band, inherently making it more difficult to detect, and likely biasing low the flare intensity after the final fits.
While we can't assume polarization characteristics in a blind search, we can flag likely polarized sources in our detection if the distribution of flagged detectors across the focal plane is representative of a highly polarized source (consistent orientation of polarized detectors flagged).

Candidate sources with fewer than five flagged detectors per frequency are typically isolated noise fluctuations and are rejected (blue markers in the figure).
This minimum also imposes a constraint on the spatial and temporal extent: in the most compact case, five detectors could correspond to just three distinct pixels across the focal plane (two overlapping pairs and one individual detector in a single band).
For example, in the 93\,GHz band with an azimuthal scan rate of $1\,\mathrm{deg}\,\mathrm{s}^{-1}$ and a sampling rate of 220\,Hz, it would take a source $0.079$\,s to traverse three separate pixels.
This implies the absolute fastest signal variation we could measure, passing our criteria, would be a signal lasting $>79$\,ms.

For each candidate source scan, we estimate its sky position using the mean RA and Dec of the clustered samples.
In practice, many detections will correspond to known, static or slowly varying sources such as bright AGN, which SO expects to detect with counts of $\mathcal{O}(10,000)$ over a 5-year survey and 40\% sky coverage.
In real-data application of our pipeline, we plan to reject scan events that occur within one beam width ($\sigma_\mathrm{FWHM}$) of cataloged AGN, to avoid saturating our detections.
However, as our simulated observations do not include known sources, this masking is not applied here.
Separate procedures are in development for analyses directed at time-domain variability in a variety of AGN species such as blazars; see, e.g., \cite{Abbasi2025} and \cite{Hincks2025}.

To refine the localization and amplitude measurement of each candidate source scan, we extract all samples within $\pm0.5$\,s of the event's median time, and within 1 degree of its estimated sky position (while this is far larger than the beam it gives us more samples to improve fitting accuracy).
This window captures the full crossing of the source over a single wafer, including samples with lower S/N that were not included in the initial clustering but still contribute information within the beam profile.

Fitting to this larger sample set improves the accuracy and robustness of the inferred amplitude and position, especially when the initial detections include only part of the beam footprint.
We model the filtered signal amplitude at each detector position, $A(\theta_\mathrm{det})$, as

\begin{equation}\label{eq:fluxfit}
    A(\theta_{\mathrm{det}})=A_0\cdot B(\left|\theta_\mathrm{det}-\theta_\mathrm{src}\right|) + C,
\end{equation} 

\noindent
where $A_0$ is the intrinsic amplitude during the scan, $\theta_\mathrm{src}$ is the best-fit source position, $B(\theta)$ is our beam profile, and $C$ is a constant offset accounting for nonzero baseline \citep{Nerval2025}. 
This model assumes the amplitude remains approximately constant over the short duration of a source crossing a wafer in a single scan.

We perform the fit using non-linear least squares optimization with \texttt{scipy} \citep{2020SciPy-NMeth} curve fit, which provides both the best-fit parameters and corresponding Jacobian.
The covariance matrix is estimated from the Jacobian where the diagonal elements yield our estimated uncertainties:  $\sigma_A$, $\sigma_\mathrm{RA}$, $\sigma_\mathrm{Dec}$ and $\sigma_C$.
As a final cut, we require that the fitted amplitude satisfies $A/\sigma_A\ge5$ to ensure high-confidence detections.

\section{Results}
\label{sec:forecasts}

\begin{figure*}
\centering
\includegraphics[width=\textwidth]{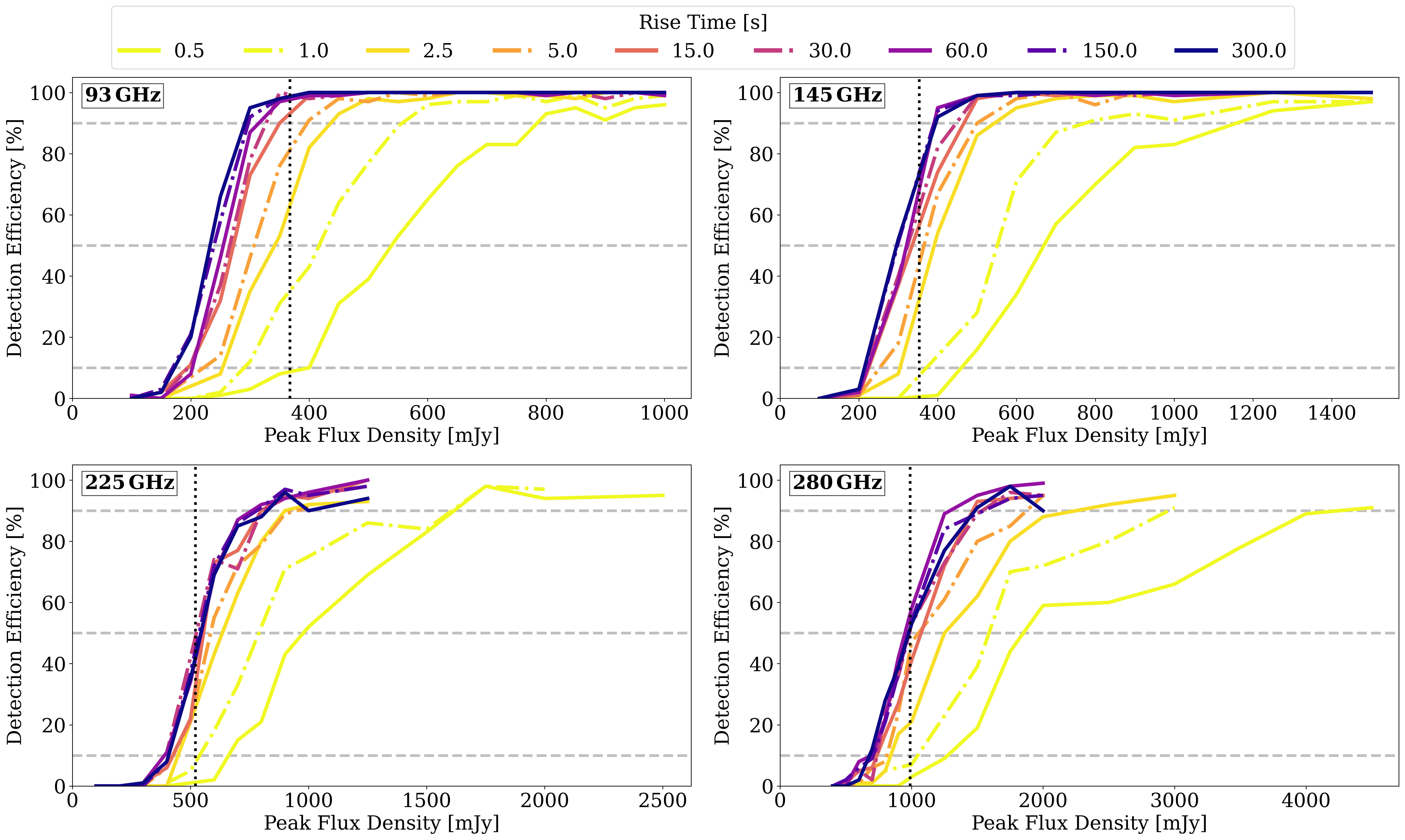}
\caption{
Detection efficiency of injected transient events at 93, 145, 225 and 280\,GHz for increasing peak flux density values and characteristic flare timescales. 
The horizontal dashed lines indicate the 10\%, 50\%, and 90\% threshold for minimal to reliably consistent detection efficiency. 
The vertical dotted lines indicate the rms noise across the time-ordered data for all wafers at each frequency. 
All frequencies show similar behavior in the increase of detection efficiency with flux density between the fastest and slowest flares. 
The main differences arise from the increased noise level at higher frequencies and increased separation between scans as the beam shrinks. 
This is particularly clear at 280\,GHz where we do not begin to cross 90\%, for any timescale, until $>$\,1\,Jy.
Note, detection efficiency here is defined as the fraction of injected flares for which a peak is detected by the full pipeline, matching in time and coordinates to the injected event.
\label{fig:detection_efficiency_f090}}
\end{figure*}

In this section we assess the performance of our SO LAT time-ordered data pipeline across a large set of simulated flare amplitudes and timescales.
We first examine the detection efficiency as a function of peak flux density and flare duration, then characterize how well the pipeline recovers the peak flux density and source position, and finally evaluate the false-positive rate.


\subsection{Flare Parameters and Recovery}
\label{subsec:params}

The light curve toy model introduced in Section~\ref{subsec:sims} allows us to generate a range of flaring events by varying two key parameters: the peak flux density, $S_p$, and the characteristic rise time, $t_r$.
While real transient light curves are rarely symmetric \citep{MacGregor2018,Biermann2024}, for controlled pipeline characterization we use symmetric decay, $t_d = t_r$.
We use this model to probe the sensitivity of our time-domain detection pipeline, particularly for rapid transients that conventional map-based approaches are likely to miss.

We focus on events with rise times from 0.5 to 300\,s, chosen to span the gap between the capabilities of time-domain and map-based methods.

The minimum of $t_r=0.5$\,s corresponds to the typical transit time of a source across a single wafer, conservatively above the 79\,ms minimum set by detector sampling (Section~\ref{subsec:cluster_cuts}).
While $500\,\mathrm{ms}\gg79\,\mathrm{ms}$, any source evolving on timescales less than 500\,ms will be isolated to a single wafer and single-directional scan of the instrument.
As the timescale decreases, the number of possible detectors hit will drop, along with the S/N and overall ability to classify.
While they may still be detectable if very bright, a minimum of $t_r=0.5$\,s allows for generally reliable fits and possible multi-wafer coverage at the faster end of transients we expect to see with SO (e.g., flaring stars).

Further, when considering speeds towards the 79\,ms minimum or lower, we may become impacted by the detector time constants $\tau_{\mathrm{eff}}$. 
If $t_r$ approaches $\tau_{\mathrm{eff}}$ we would expect to see stronger S/N suppression.
Early characterization of SO LAT detectors suggests typical time constants in the MF channels should be around 1--5\,ms, far below the values of $t_r$ where we would expect to have sources survive our cuts \citep{Dutcher2024}. 

The maximum, $t_r=300$\,s, approaches the time scale where map-based detection becomes more effective (typically $\sim5$--$10$\,minutes) for initial detection. But map-based detection still struggles to resolve substructure in time.
In total, we sample across $t_r=$ 0.5, 1, 2.5, 5, 15, 30, 150, 300\,s.
Future work may investigate shorter timescales approaching the 79\,ms limit, which will become feasible once the initial science observations are complete and SO finalizes its scan speeds, focal plane configuration and scan strategy.
ACT, for example, has explored much shorter cadences of $2.5$\,ms in a dedicated study of the Crab Nebula pulsar \citep{Guan2025}.

We concentrate our pipeline testing on the MF (93 \& 145\,GHz) bands where both sensitivity and control of synchrotron foreground emission are optimal (see Section~\ref{subsec:SO}). UHF (225 \& 280\,GHz) bands are also explored, primarily to assess detection at fast timescales rather than faint flux densities.
The LF (27 \& 39\,GHz) bands have not been included, as we expect these channels to have substantially poorer blind transient detection capability than the MF bands due to their significantly larger beam sizes and much smaller detector counts. 
Although the larger LF field of view increases the instantaneous sky coverage, we expect this advantage to be outweighed by the reduced point-source sensitivity and localization capability, particularly for the rapid transients considered here. 
The LF channels are instead expected to be particularly valuable for providing complementary low-frequency measurements of transients identified in higher-frequency bands.

We choose flux density peaks in our simulations as follows:

\begin{itemize}
    \item \textbf{93\,GHz (MF):}\\
        \hspace*{1.2em}100--1000\,mJy \dotfill (increments of 50\,mJy)
    \item \textbf{145\,GHz (MF):}\\ 
        \hspace*{1.2em}100--1000\,mJy \dotfill (increments of 100\,mJy)\\
        \hspace*{1.2em}1--1.5\,Jy \dotfill (increments of 250\,mJy)
    \item \textbf{225\,GHz (UHF):}\\
        \hspace*{1.2em}100--1000\,mJy \dotfill (increments of 100\,mJy)\\
        \hspace*{1.2em}1--2.5\,Jy \dotfill (increments of 250\,mJy)
    \item \textbf{280\,GHz (UHF):}\\
        \hspace*{1.2em}400--1000\,mJy \dotfill (increments of 100\,mJy)\\
        \hspace*{1.2em}1--2\,Jy \dotfill (increments of 250\,mJy)\\
        \hspace*{1.2em}2--4.5\,Jy \dotfill (increments of 500\,mJy)
\end{itemize}

Expecting the 93\,GHz band to be the most sensitive, we opted for the smallest increments but increased from 50\,mJy to 100\,mJy gaps at higher frequencies to save on computation time when running the simulations for each parameter combination.
The maximum peak flux density injected is increased from 1000\,mJy at 93\,GHz to 4.5\,Jy at 280\,GHz to ensure that the curve for each value of $t_r$ is able to cross the 90\% detection efficiency threshold.
This accounts for the typical noise rms increasing in higher frequency bands.

\begin{deluxetable*}{l|ccc|ccc|ccc|ccc}
\tablecaption{Flux Density Detected at 10\%, 50\%, and 90\% Efficiency}
\tablewidth{0pt}
\tablehead{
$t_r$ (s) & \multicolumn{3}{c|}{93\,GHz (mJy)} & \multicolumn{3}{c|}{145\,GHz (mJy)} & \multicolumn{3}{c|}{225\,GHz (mJy)} & \multicolumn{3}{c}{280\,GHz (mJy)} \\
 & 10\% & 50\% & 90\% & 10\% & 50\% & 90\% & 10\% & 50\% & 90\% & 10\% & 50\% & 90\%
 }
\startdata
0.5   & 400  & 550  & 800  & 450  & 650  & 1150 & 650   & 1000 & 1650  & 1250  & 1900 & 4250 \\
5     & 250  & 300  & 400  & 250  & 350  & 500  & 425   & 600  & 900   & 800   & 1100 & 1900 \\
30    & 200  & 275  & 325  & 225  & 350  & 450  & 400   & 550  & 800   & 750   & 1000  & 1500 \\
150   & 175  & 225  & 300  & 225  & 300  & 400  & 400   & 550  & 750   & 700   & 1000 & 1600
\enddata
\tablenotetext{}{
Flux density thresholds at 10\%, 50\%, and 90\% detection efficiency for various flare durations, $t_r$, and frequency bands.
Note the sensitivity to recovered events is significantly improved between $t_r<1$\,s and $t_r<10$\,s in all bands but converges towards the typical rms of time-ordered data for $t_r>10$\,s.
}
\label{tab:results}
\end{deluxetable*}
 
For each combination of $t_r$ and $S_p$, 100 simulated light curves are injected into the time-ordered data at randomly sampled peak times and spatial coordinates.
Our pipeline is then run to blindly detect the injected events and estimate detection rates.
The injection rate is deliberately non-physical, as the aim is to probe detection efficiency across a broad parameter space.
Observational constraints on transient rates at millimeter-wavelengths -- especially for variability on timescales $\lesssim\mathcal{O}(\mathrm{days})$ -- remain few and far between.

Throughout this section, a detection is counted if an injected flare is matched to the detected event to within 0.5\,s of its peak time and the position is within $0.5^\prime$. 
The outputs from our full detection pipeline can then be directly compared to their matching injected flares.
Any false detections are not counted as detections in this context, so our efficiency curves reflect the true recovery rate of injected sources, not the rate of arbitrary detections (false detections are explored in Section~\ref{subsec:false_detections}). 

\begin{figure}
    \centering
    \includegraphics[width=\linewidth]{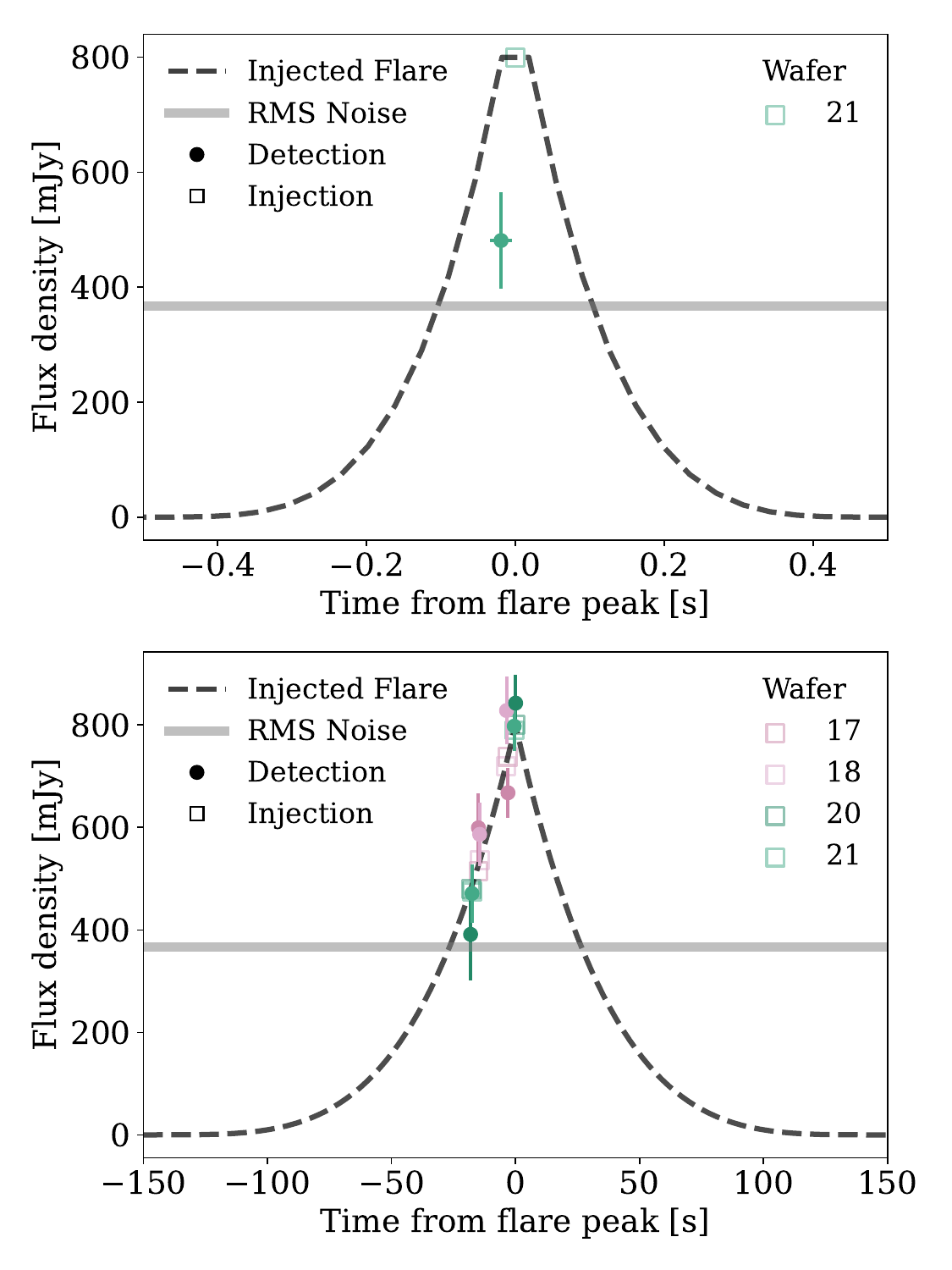}
    \caption{
    Examples of simulated flares (black dashed line) with different durations (top: 0.5\,s to peak, bottom: 150\,s to peak), and peaks detected in the 93\,GHz band by our method (colored circles). 
    The squares represent points along the light curves when the source crossed the respective wafer in the simulated observations.
    The gray horizontal bar corresponds to the rms noise of time-ordered data in the 93\,GHz wafers ($367\pm4$\,mJy).
    For the rapid $t_r=0.5$\,s flare, as expected, the observation only scans over the source once, and only across a single wafer.
    We are still able to make a detection that passes our criteria; however, we observe a significant underestimation of the peak flux density.
    The slower, $t_r=150$\,s flare, is easily detectable in each of its scans, being observed to pass twice across the entire focal plane (seen by four wafers in one direction and then by the same four in the opposite direction).
    }
    \label{fig:detected_lightcurves}
\end{figure}

\subsection{Detection Rates and Limits}
\label{subsec:rates}

The full detection efficiency curves for our injected flares in the simulated SO LAT MF and UHF observations are shown in Figure~\ref{fig:detection_efficiency_f090}.
We achieve high completeness ($\ge90$\% detection efficiency) for 0.5\,s transients with peak flux densities of 800\,mJy at 93\,GHz, 1150\,mJy at 145\,GHz, 1650\,mJy at 225\,GHz, and 4.25\,Jy at 280\,GHz. 
Detection efficiency improves substantially for longer-duration flares: for $t_r\ge30$\,s the 90\% detection efficiency thresholds drop to 325\,mJy, 450\,mJy, 800\,mJy, and 1.5\,Jy at the respective frequency bands, as shown in Table~\ref{tab:results}.

The 50\% detection level is reached for 0.5\,s flares at 550, 650, 1000 and 1900\,mJy for 93, 145, 225, and 280\,GHz respectively.
At the 10\% threshold, detectable flux densities drop further across bands to 400, 450, 650, and 1250\,mJy respectively.
Notably, we achieve $\gtrsim50$\% detection efficiency in all frequencies for flares rising slower than 5\,s at flux densities near or below the typical rms of time-ordered data.
This can be seen in the 50\% detection efficiency column in Table~\ref{tab:results}, where the reported flux densities for flares with $t_r\ge5$\,s are consistent with the reported rms for all frequencies except 93\,GHz.
The 93\,GHz channel performs better due to a combination of lower noise and more pixel overlap with the slightly larger beam.

The flux densities crossing the 10\% detection threshold show very little variation for $t_r\ge5$\,s, decreasing by no more than 100\,mJy across all frequencies as flares increase in duration. 
For $t_r\ge30$\,s the difference is at most 50\,mJy.
This suggests we are reaching a critical noise limit in time-ordered data, where increasing flare duration no longer significantly improves detection efficiency.

The overall behavior we see in Figure~\ref{fig:detection_efficiency_f090} across the four frequency bands is consistent, with a clear shift of the detection efficiency curves towards higher peak flux densities at increasing frequencies.
This arises from two main factors: noise scaling and detector response.
The rms noise in time-ordered data is relatively consistent for our MF frequencies with $367\pm4$\,mJy at 93\,GHz and $353\pm4$\,mJy at 145\,GHz, but rises quickly in the UHF bands to $520\pm5$\,mJy at 225\,GHz and $988\pm10$\,mJy at 280\,GHz (vertical dotted lines in Figure~\ref{fig:detection_efficiency_f090}).
This matches our expectation from Section~\ref{subsec:survey_strat} that the best sensitivity to sources will be observed at 93 and 145\,GHz.

Despite having matching physical wafer layouts between bands, the UHF pixels are less effective in this context than MF bands.
The 93\,GHz channel has instantaneous sky coverage $4.81\times$ and $5.95\times$ larger than the 225\,GHz and 280\,GHz channels, respectively.
This geometric factor, along with higher rms noise, translates to proportionally fewer transient detections in UHF bands.
As the beam size decreases, it becomes increasingly difficult to obtain confident samples around a source as we are likely to miss much of the variation, particularly for the fastest sources.
The higher peak flux densities required to cross the 90\% detection threshold at UHF frequencies demonstrate this effect.

We observe a sharp rise in detection efficiency over a few hundred mJy for all flares with $t_r\ge15$\,s, reflecting high sensitivity to signals varying at timescales of tens of seconds or more. 
In contrast, flares shorter than $\sim5$\,s show flatter efficiency curves, with detection rates increasing more slowly as the rise time shortens.

The flattened slopes for fast events arise from poor S/N when flare durations approach the sampling cadence.
Shorter flares are more impulsive, with their signal received over a very brief time window.
When the flare duration falls below timescales of instrumental drift or atmospheric variation (seconds), the signal can be diluted or confused with noise fluctuations.
As a result, increasing the peak flux density only moderately improves detection odds, leading to the flattened slopes seen in Figure~\ref{fig:detection_efficiency_f090}.

\begin{figure}
    \centering
    \includegraphics[width=\linewidth]{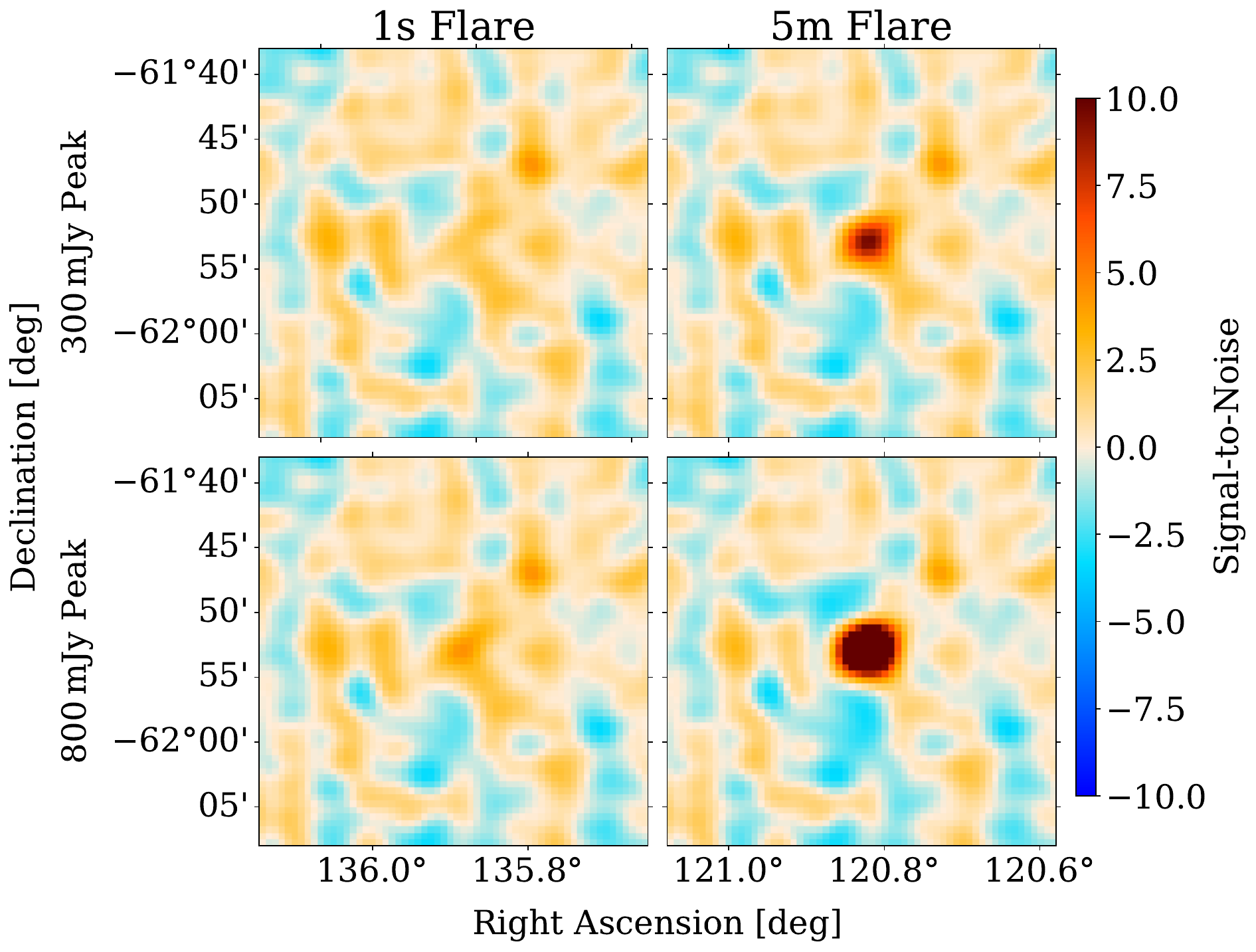}
    \caption{
    Matched filter map cutouts of simulated flares injected into time-ordered data with amplitudes of 300 and 800\,mJy (top and bottom rows respectively) and for flare durations of 1\,second and 5~minutes (left to right). 
    These thumbnails demonstrate where the two methods meet and why both are required for different cases.
    For example, a very bright 800\,mJy source begins to appear in the matched-filter response for a rapid flare but remains below the minimum S/N\,$\ge5$, comparable to noise fluctuations seen nearby. With our blind time-domain search we can recover this same source with $\ge90$\% detection efficiency. 
    }
    \label{fig:d1_comparison}
\end{figure}

\begin{figure*}
\centering
\includegraphics[width=\textwidth]{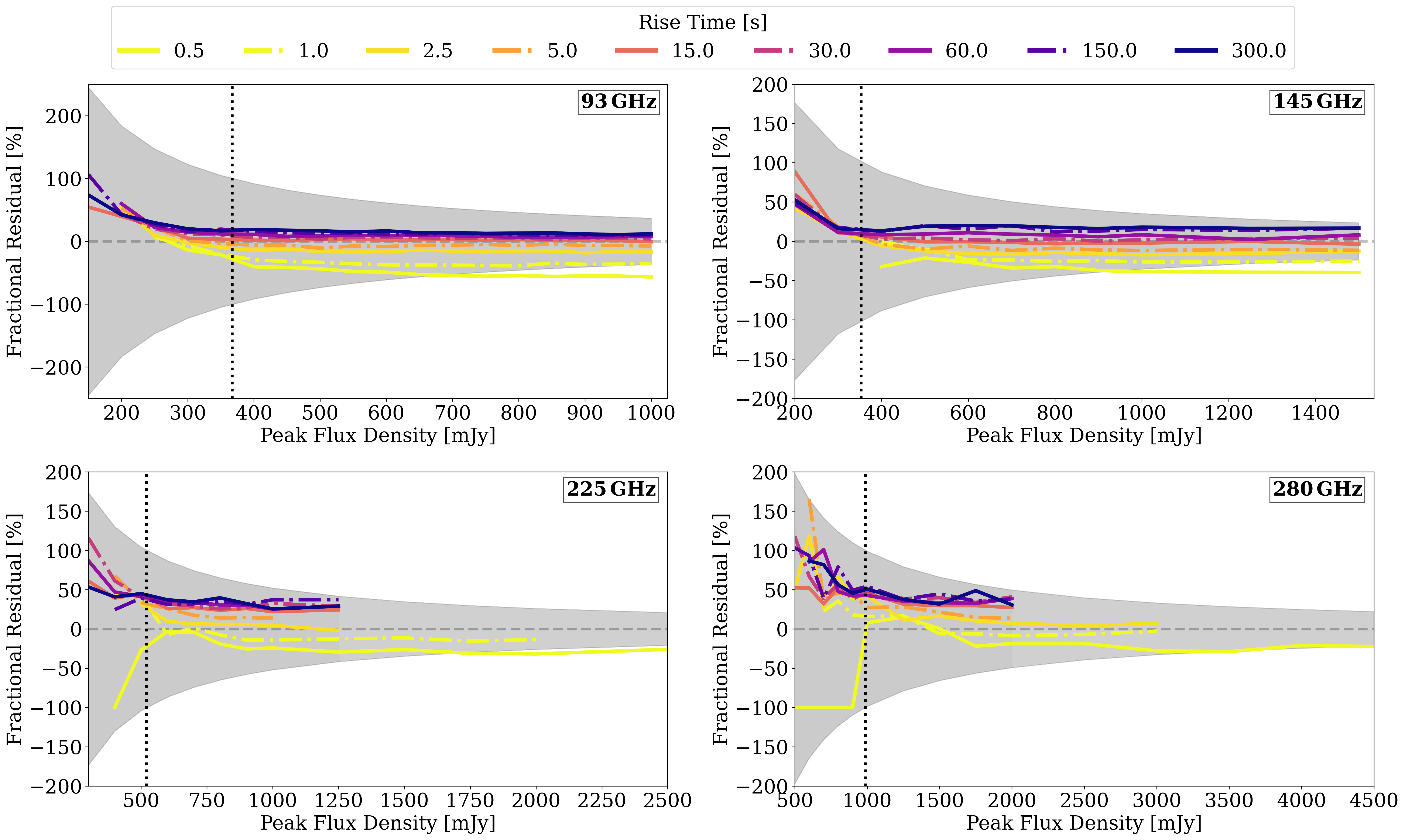}
\caption{
        Fractional peak flux density residuals for each parameter combination at 93, 145, 225 and 280\,GHz. The vertical dotted lines indicate the mean rms across time-ordered data, for all wafers at each frequency. The gray shaded region indicates the rms / peak flux density to show the expected noise bias. As with the detection efficiency, each band shows consistent behavior where we see the flux density for slower flares, $t_p\ge15$\,s, is well constrained within the expected noise bias. 
        For fast flares we see an underestimation of peak flux density and an overestimation for the slowest flares. Note, however, the sharp upturn to an overestimation of $\gtrsim50$\% at the lowest peak flux densities; this is because the flux density is close to the typical rms of time-ordered data, thus limiting the accuracy of the amplitude fit.
\label{fig:flux_residuals}
}
\end{figure*}

Figure~\ref{fig:detected_lightcurves} shows representative flare recoveries at 93\,GHz using our pipeline based on time-ordered data.
For the fastest flare ($t_r=0.5$\,s, top panel), only a single wafer samples the source during its lifetime.
It is correctly detected at the expected time and wafer, consistent with the $\ge90$\% detection efficiency for an 800\,mJy peak over this timescale, although the measured flux density is underestimated as only a few detectors sample the peak (see Section~\ref{subsec:flux_residuals}).
The slower flare ($t_r=150$\,s, bottom panel) is detected in multiple wafers across successive scans, with measured flux densities close to the true light curve values.
Even so, this scanning pattern means only about 50\,s of the flare is observed.
This highlights the role of sky position in observational completeness.

Our time-domain approach provides sensitivity to rapid transients that map-based methods cannot detect.
To demonstrate this further, we produced matched-filtered single-wafer maps for injected flares with durations of 1\,second ($t_r=t_d=0.5$\,s) and 5\,minutes ($t_r=t_d=150$\,s), and peak fluxes of $300$\,mJy and $800$\,mJy (Figure~\ref{fig:d1_comparison}).
These maps integrate $\sim$3--5\,minutes of data per wafer and represent the fastest cadence achievable within standard map-domain pipelines.

As noted in Section~\ref{subsec:SO}, the fully populated SO LAT receiver will have a baseline single-wafer sensitivity at 93\,GHz of 46.5\,mJy with $\sim3$\,minutes of map integration over a source.
Rapid events, however, are strongly diluted in map-making.
The 300\,mJy flare, with $t_r=0.5$\,s (far below the integration-time) is completely washed out, while even the 800\,mJy flare is suppressed to $\mathrm{S/N}=4.19$, below the $\mathrm{S/N}\ge5$ blind-detection threshold for map-based searches.
By contrast, our time-domain pipeline recovers this 800\,mJy flare with $\ge90$\% detection efficiency as shown in Figure~\ref{fig:detected_lightcurves}.

Longer events are easily detectable in both approaches, but because their duration here is comparable to the map cadence, they would likely be absent in maps taken immediately before or after the peak.
While we may be able to observe a source flaring in a single map, it can be difficult to classify or confidently claim variability with a single data point, compared to multiple scans detected in the time-domain showing a clear rise and/or fall.
This illustrates a key advantage of the time-domain approach: sensitivity to short flares that are missed in maps, and the ability to capture multiple scans of a single event for improved characterization. 

\subsection{Flux Density Recovery}
\label{subsec:flux_residuals}

We now assess how accurately we recover the peak flux density of detected flares.
Here, we compare the fitted peak amplitude of sources, made near the true peak time of an injected flare, to the injected peak flux density. 
We focus on recovering the peak flux density rather than fitting the full transient light curve, as this allows us to assess the accuracy of individual detections across scans or wafers.
These could then be combined to characterize variability and construct light curves of detected sources.

For flares with $t_r\ge5$\,s, we recover peak flux densities to within $\pm10-20$\% of the injected value in the MF bands and $\pm30-40$\% in the UHF bands, consistent with expected noise bias.
Figure~\ref{fig:flux_residuals} shows the fractional peak flux density residuals for all detections across the 93, 145, 225, and 280\,GHz bands.
As with detection efficiency, consistent behavior can be seen across frequencies, with deviations becoming magnified in the UHF bands, particularly at 280\,GHz, where sources are typically under-sampled and contaminated with higher rms noise levels.

The preference for overestimation seen in slower sources ($t_r\ge150$\,s) in MF bands can be attributed to the flux drifting noticeably across the scan window.
Equation~\ref{eq:fluxfit} assumes the amplitude $A_0$ is constant during the scan.
For slow flares, drift across the window of samples used in the fit may cause it to assign part of the long-timescale trend to the beam amplitude $B$, rather than offset from the baseline $C$.
This injects a positive bias into $A_0$ and overestimates the results.

Ultimately, these percentage offsets equate to $\pm50$--$100$\,mJy for most sources, within the bands of the expected noise bias for all cases, save the fastest events ($t_r=0.5$\,s) which we discuss below.
These are acceptable margins for initial detections.
All sources detected through this pipeline will require follow-up analysis where we can run stronger fits on now known coordinates and timestamps, with more dedicated noise modeling and rigorous signal calibration. 

For rapid events with rise times $<5$\,s, we observe a systematic underestimation of the peak flux density.
This bias is a consequence of limited detector sampling: a flare rising to peak within $\sim0.5$\,s would cross only a few detectors at its peak brightness, if not missing the peak entirely in the sampling window.
As a result, the fit lacks sufficient information to recover the true peak.
We remind the reader that we set $t_r=t_d$ in Section~\ref{subsec:params}.
Consequently, if we were to either further decrease both, or relax one side for these fastest flares, the sampling will be sparser and recovering the flux density will be even more difficult. 

It's also worth noting an upturn and systematic overestimation in the recovered peak flux density at the lowest injected amplitudes, across both MF and UHF bands. 
These flux densities are at or below the typical rms of time-ordered data, meaning the signal is no longer clearly distinguishable from noise.
In these cases, the peak identified as the initial guess for fitting may correspond to a random noise fluctuation rather than the true flare maximum at that time, leading the fit to overestimate the flux density closer to the rms of time-ordered data itself (the residuals are on average 50--100\% for peak flux densities injected at $\sim$half the typical rms of time-ordered data).

This is consistent with the minimum observable flux densities for slow flares shown in Table~\ref{tab:results}, and highlights the limitations of amplitude recovery near the noise floor.
In contrast, the underestimation and increased scatter seen for the fastest flares ($t_r=0.5$\,s) at 225 and 280\,GHz likely reflects poor sampling -- these events span too few detectors near their peak to enable a reliable fit, especially near our detection threshold.

To mitigate noise limitations, post-detection analysis of sources detected with our method can be performed.
This would involve using, for example, forced photometry on the known position and times of the flare, which is far more sensitive but more computationally expensive and requires prior knowledge of the source.

\subsection{Pointing Recovery}
\label{subsec:position_residuals}

\begin{figure}
    \centering
    \includegraphics[width=\linewidth]{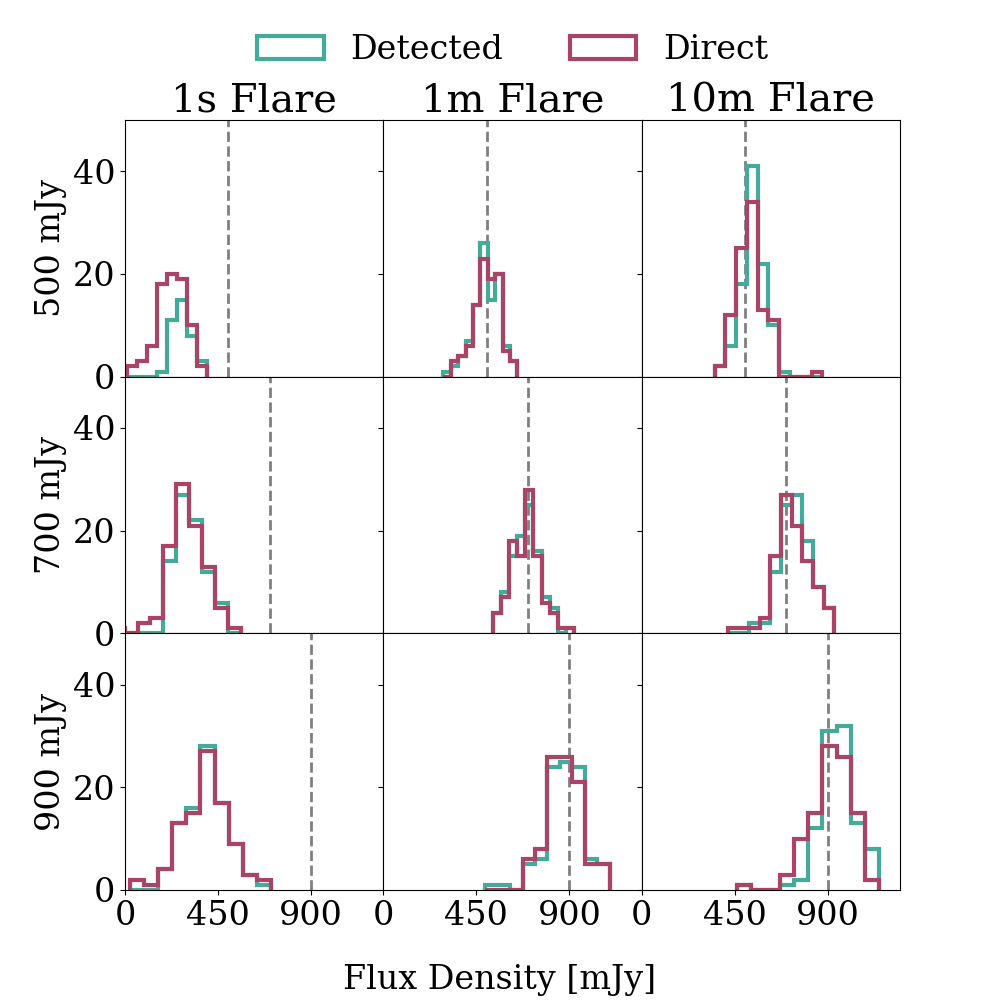}
    \caption{
        Histograms of the peak flux density for all flares in a given search in the 93\,GHz band, measured by fitting at the known injected source location (direct: magenta) and at the blindly estimated locations following the detection pipeline (detected: teal). The histograms are shown for flare durations (columns) of 1\,second, 1\,minute and 10\,minutes, and at peak flux density values (rows) of 500, 700 and 900\,mJy. We see well recovered flux density for the 1 and 10\,minute flares, and even with a consistent underestimation from the injected peak flux density, the 1\,second flares are still centered well around measurements from fitting the flux density directly at the injected source positions.
        This indicates the offset is an observational constraint where the source flares are too fast for the detectors to fully capture.  
    }
    \label{fig:flux_histograms}
\end{figure}

\begin{figure*}
    \centering
    \begin{tabular}{cc}
        \includegraphics[width=0.45\textwidth]{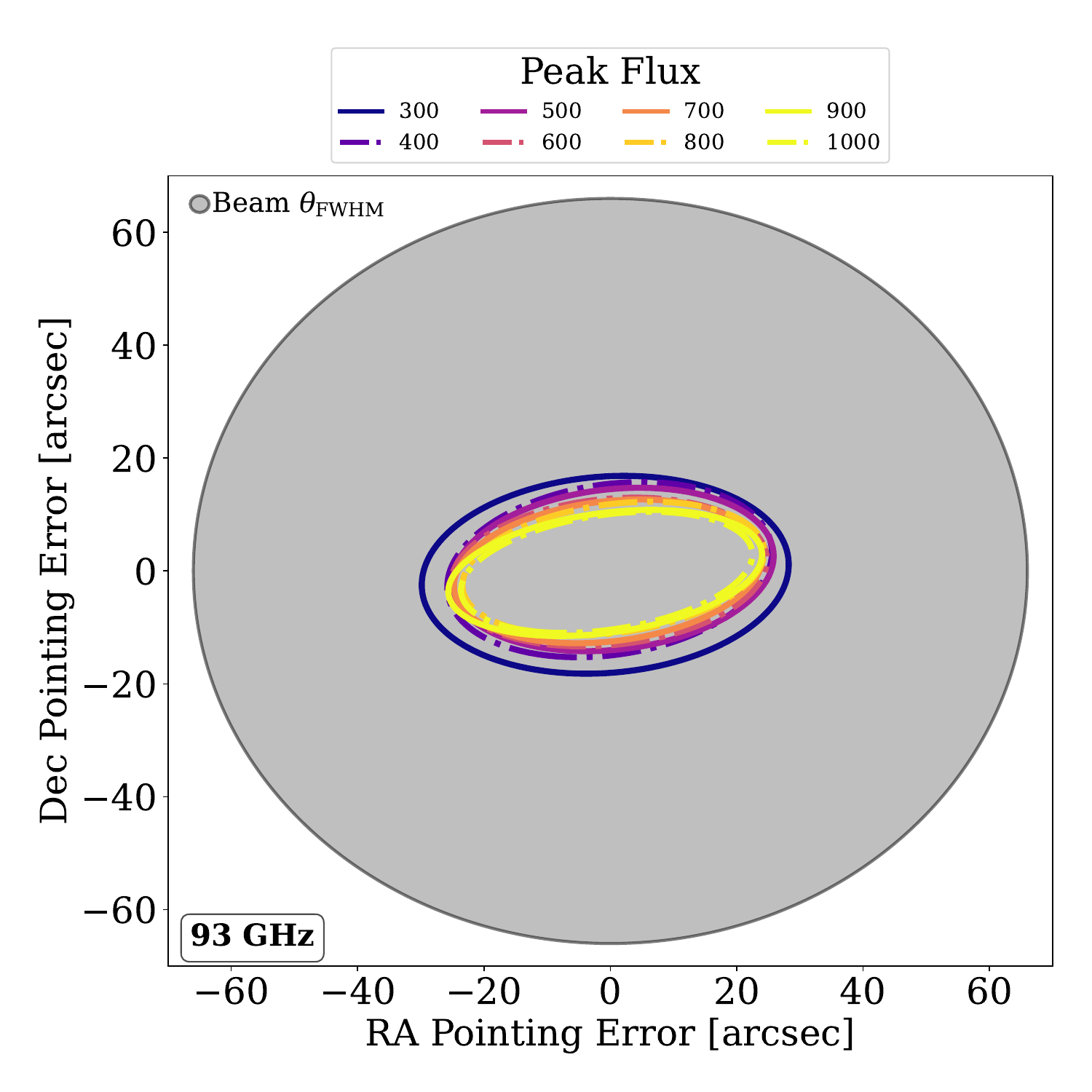} &
        \includegraphics[width=0.45\textwidth]{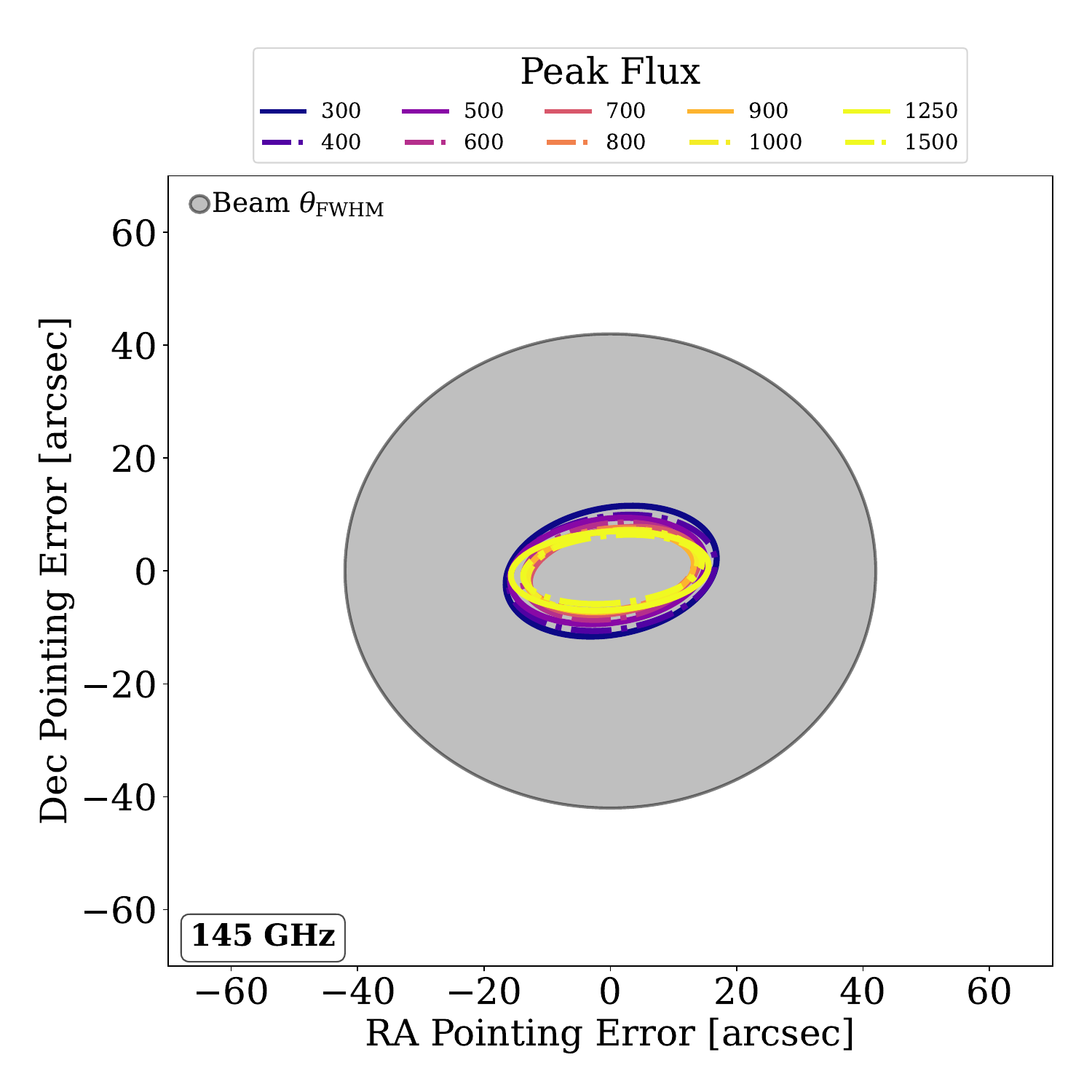} \\
        \includegraphics[width=0.45\textwidth]{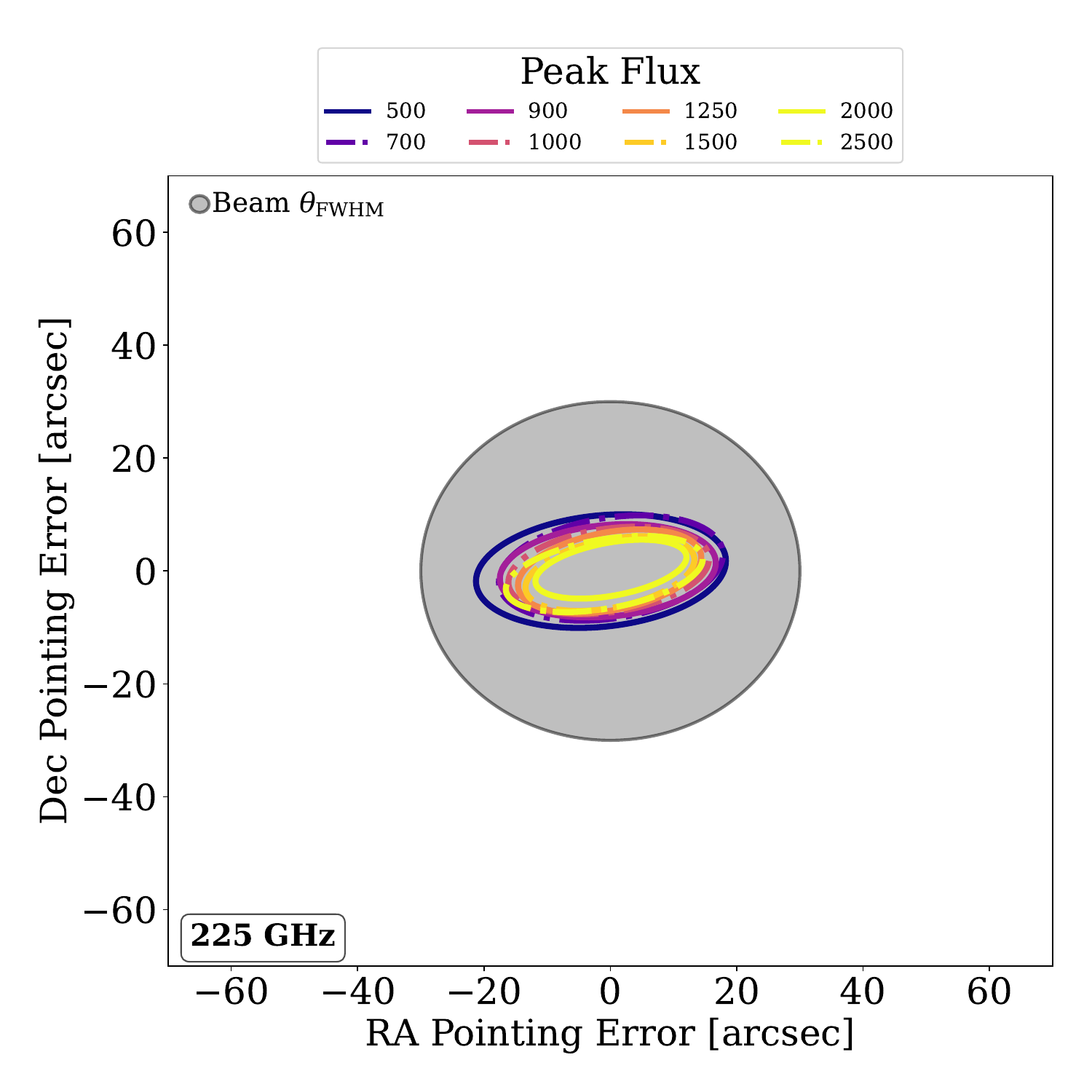} &
        \includegraphics[width=0.45\textwidth]{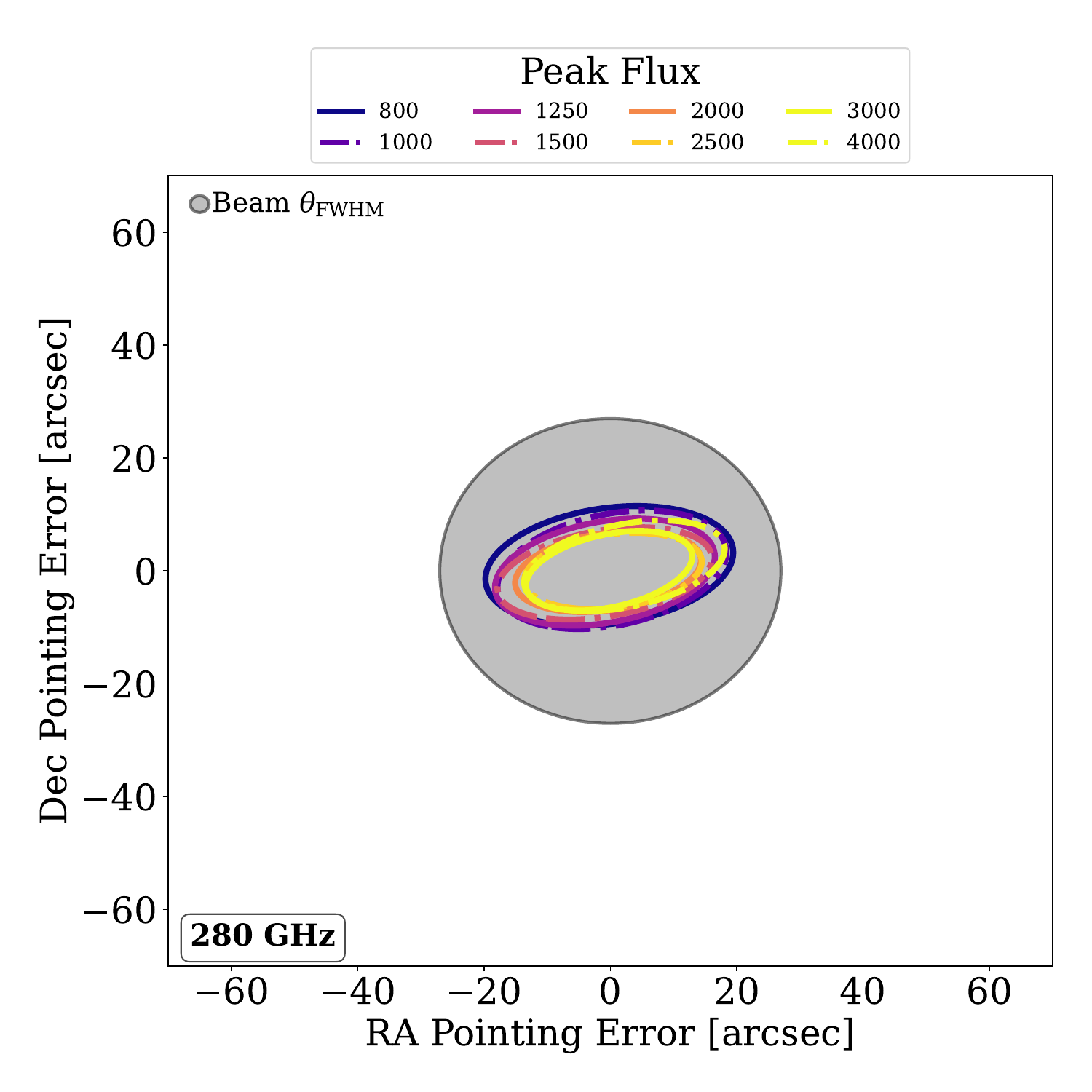} \\
    \end{tabular}
    \caption{
    Contours of the 1$\sigma$ error in the fit source positions for increasing flare peak flux density observed at 93, 145, 225 and 280\,GHz. The gray shaded area shows the beam FWHM ($\theta_\mathrm{FWHM}$). We are able to constrain the source location to within 8--17\% of the beam size at each frequency (Table~\ref{tab:sens}).
    }
    \label{fig:pointing_error}
\end{figure*}

While we have demonstrated well constrained recovery of the peak flare intensity, we consider if potential pointing offsets could further explain the fractional flux density offsets we do observe, particularly for the rapid sources where we see consistent underestimation in Figure~\ref{fig:flux_residuals}.
We compare our flux density fit at the locations blindly estimated through our pipeline to a fit directly on the known injected source locations for a subset of parameter combinations.
The results of this are shown in Figure~\ref{fig:flux_histograms} where we have the detected vs directly fit flux density for 100 simulated flares with durations of 1\,second, 1\,minute and 10\,minutes, and injected flux densities 500, 700, and 900\,mJy.

For all combinations, flux densities recovered for the source positions estimated through our pipeline are comparable to the true distributions. 
The flux densities recovered for the 1 and 10\,minute duration flares fall within the expectation of the true injected peaks (dashed black line).
For the fast events we clearly see an underestimation of both distributions from the injected value.
This is consistent with our previous comments on the flux density offsets arising from observational and noise limitations rather than a modeling error or imprecision in pointing.

Minimizing any pointing offset is crucial not just for the flux density fits, but also to enable accurate follow-up observations.
Figure~\ref{fig:pointing_error} shows how well we are able to constrain the true source location.
Across all frequency bands, we achieve a median offset of 9.8$^{\prime\prime}$, with 68\% of sources localized within 13.4$^{\prime\prime}$ of their true positions.
The median pointing precision of 14.6$^{\prime\prime}$, 7.0$^{\prime\prime}$, 9.4$^{\prime\prime}$, and 9.1$^{\prime\prime}$ at 93, 145, 225, and 280\,GHz respectively, corresponds to 8-11\% of the beam FWHM at each frequency.

The errors in source position are strongly constrained for all MF and UHF bands, where $1\sigma$ contours are comparable to $\sim\sigma_\mathrm{FWHM}$ of the beam at each frequency (with the beam FWHM shown as the gray circles).
This demonstrates that, for even the fastest sources with few detections, the average source position is well constrained.

Time-ordered data positional accuracy is expected to be comparable with map-based localization as both are derived from the same underlying pointing calibration and beam response. 
Mapmaking does not intrinsically improve the pointing accuracy, instead increasing the S/N through the combination of repeated measurements and making the source position easier to estimate. 
This is the argument used in the construction of pointing corrections for both ACT DR4 and DR6 \citep{Aiola2020, Naess2025}, with map-based corrections preferred for efficiency rather than accuracy. 
Therefore, after accounting for differences in effective S/N and sampling effects, the positional uncertainties inferred from time-ordered data and maps should in principle be consistent.

\subsection{False Detections}
\label{subsec:false_detections}

Finally, we estimated the rate of false positives contaminating our detections by comparing detected and injected sources in our simulations.
After running each observation many times, we found an average of four false positives across our two hours of observation in the MF channels, and two in UHF channels.
If we consider, for consistency, the cases when we are able to recover all of our injected sources, this equates to a 95\% true detection and a 5\% false detection rate of sources.

In absolute terms, this yields $\sim$2 ($\sim1$) false positives per hour in MF (UHF) channels -- equivalent to $\sim20-50$ false detections per day when scaling up to a full day of observations across all optics tubes.
Whilst these counts seem significant, nearly all false detections lie close to the minimum thresholds required to pass our clustering and filtering criteria (Section~\ref{subsec:cluster_cuts}), suggesting that most can be identified and excluded through additional scrutiny.
Furthermore, because the noise background in these simulations is fixed, the false positive rate remains relatively stable across different injection parameters.

To mitigate false positives, we would cross-check low-confidence detections through multi-frequency coincidence, forced photometry at the candidate location, and temporal consistency checks.
These steps are particularly valuable during early science operations, where instrumental glitches that survive early versions of the data cuts may mimic transient signals.

It is worth noting that if we have a true rate of one transient event per day, or even one per week, then 4 in two hours would be a large rate. 
However, without accurate modeling of the systematics and noise from the SO LAT in full operation we cannot fully contextualize this.
Refined cuts and post-detection validation will be crucial for balancing the true data detection efficiency and false positive rate.
Overall, our findings demonstrate that the false positive rate is low and well-characterized in our simulated observations.

\section{Conclusion}
\label{sec:conclusion}

Our time-domain transient pipeline enables the blind detection of rapidly varying astrophysical sources directly within SO LAT time-ordered data, accessing a parameter space largely unexplored by conventional map-based searches.
We achieve $\ge90$\% detection efficiency for sub-second flares at peak flux densities of 800\,mJy at 93\,GHz and 1150\,mJy at 145\,GHz.
These events would be completely undetectable or fall below detection thresholds in the highest cadence maps.
For longer-duration transients (tens of seconds to minutes), we recover the majority of events at flux densities near the typical RMS of time-ordered data, recovering peak amplitudes to within 20\% and localizing sources to within 8-11\% of $\theta_{\mathrm{FWHM}}$ across all MF and UHF bands.

By targeting timescales spanning 0.5 to 300\,s, from the regime below map-based detection capabilities towards the limits of time-domain sampling, we demonstrate that time-domain analysis of CMB survey data can effectively probe a previously inaccessible regime of millimeter-wave variability.
Our pipeline and map-based approaches are naturally complementary: time-domain detections can boost low-confidence map-based candidates, resolve sub-integration variability, and provide precise timestamps for multi-wavelength coordination, while maps contribute sensitivity to fainter, slower sources.
Together, these methods position SO to build comprehensive millimeter-wave transient catalogs spanning a wide range of timescales and flux densities.

The ability to detect sources across multiple scans and wafers is crucial for constructing detailed light curves, performing follow-up analyses, and enabling rapid public alerts.
Multi-frequency detections will further allow us to constrain source characteristics and emission mechanisms, distinguishing between thermal and non-thermal processes.
While we observe systematic underestimation of peak flux densities for the fastest flares ($t_r<5$\,s) due to sparse detector sampling (an effect magnified at higher frequencies), these biases are well-characterized and can be mitigated through forced photometry and improved noise modeling as SO transitions into science operations.

Characterization of real SO LAT noise properties and instrumental systematics during early observations will inform refinements to our pre-processing and detection criteria, ensuring on-sky performance approaches the results shown here.
The capabilities of the instrument will be amplified as SO expands to its full 13-tube configuration, doubling the total detector count \citep{Abitbol2025}.
The pipeline developed here provides the framework to access the rapidly varying regime of millimeter-wave transients, positioning SO to deliver improved population statistics, stronger astrophysical constraints, and potentially the discovery of new classes of millimeter-wave transients.

\section{Acknowledgments}
\label{sec:acks}
This work was supported in part by a grant from the Simons Foundation (Award \#457687, B.K.). 
This work was supported by the U.S. National Science Foundation (Award Number: 2153201).
This research used resources of the
National Energy Research Scientific Computing Center, which is
supported by the Office of Science of the U.S. Department of Energy under Contract No. DE-AC02-05CH11231.
This research was supported in part by The Dr Albert Shimmins Fund through the Albert Shimmins Postgraduate Writing Up Award (University of Melbourne). 
Melbourne authors acknowledge support from the Australian Research Council’s Discovery Projects scheme (DP210102386).
CB acknowledges partial support by the Italian Space Agency LiteBIRD Project (ASI Grants No. 2020-9-HH.0 and 2016-24-H.1-2018), and the Italian Space Agency Euclid Project, as well as the InDark and LiteBIRD Initiative of the National Institute for Nuclear Physics, and Project SPACE-IT-UP  by the Italian Space Agency and Ministry of University and Research, Contract Number  2024-5-E.0 and the RadioForegroundsPlus Project HORIZON-CL4-2023-SPACE-01, GA 101135036, and the CMB-Inflate project funded by the European Union’s Horizon 2020 Research and Innovation Staff Exchange under the Marie Skłodowska-Curie grant agreement No 101007633.
ND acknowledges funding from the European Union (ERC, POLOCALC,
101096035). Views and opinions expressed are, however, those of the authors only and do not necessarily reflect those of the EU or the ERC. Neither the EU nor the granting authority can be held responsible for them.
YG acknowledges support from the University of Toronto's Eric and Wendy Schmidt AI in Science Postdoctoral Fellowship, a program of Schmidt Sciences. The Dunlap Institute is funded through an endowment established by the David Dunlap family and the University of Toronto.
CHC acknowledges BASAL CATA FB210003.
ADH acknowledges support from the Sutton Family Chair in Science, Christianity and Cultures, from the Faculty of Arts and Science, University of Toronto, and from the Natural Sciences and Engineering Research Council of Canada (NSERC) [RGPIN-2023-05014, DGECR-2023-00180].
RH acknowledges the NSERC Discovery Grant RGPIN-2025-06483 and Arthur B. McDonald Fellowship.
KMH acknowledges NSF-AST award 2533575.
Some of the results in this paper have been derived using the \texttt{scipy}, \texttt{scikit-learn}, and \texttt{sotodlib} packages.

\bibliographystyle{aasjournal.bst}
\bibliography{main}

\end{document}